\documentclass[11pt]{article}
\usepackage{geometry}                
\usepackage[parfill]{parskip}    
\usepackage{graphicx}
\usepackage{amsmath, amsthm, amsfonts}
\usepackage{amssymb}

\usepackage{color}
\usepackage[colorlinks]{hyperref}
\hypersetup{linkcolor=blue,%
citecolor=red,%
urlcolor=cyan}
\usepackage{url} 

\usepackage{epstopdf}
\title{\bf Quantum Field-Theoretic Predictions of $\bf{\Psi}$-Epistemic Models of Quantum Mechanics}

\author { \.{I}nan\c{c} \c{S}ahin $^a$\footnote{inancsahin@ankara.edu.tr}, 
\date{}                        

\begin{document}

\maketitle

\begin{abstract}
$\Psi$-epistemic models of quantum mechanics imply that the quantum state does not correspond to physical reality, but instead reflects the observer’s knowledge of the underlying quantum system. The epistemic view of the quantum state has the potential to shed light on several foundational problems of quantum theory and has attracted considerable attention in the literature. On the other hand, the Pusey–Barrett–Rudolph theorem demonstrated that broad classes of $\psi$-epistemic models must lead to predictions that deviate from those of quantum mechanics. Although the original theorem involved entangled joint measurements on composite systems, alternative no-go theorems involving measurements on single quantum systems were developed shortly thereafter. Experimental investigations of the deviations predicted by $\psi$-epistemic models from quantum mechanics are still ongoing. So far, such tests have been performed within the framework of non-relativistic quantum mechanics and predominantly rely on quantum information based measurement procedures. In this work, we show that $\psi$-epistemic models can give rise to deviations from standard quantum field-theoretic predictions through modifications of polarized scattering cross sections and decay widths. Our results do not require a relativistic formulation of ontological models or of the Harrigan–Spekkens criterion; the essential assumption is merely that measurements implemented through relativistic processes can still be represented within the ontological framework by well-defined response functions and probabilities. The present work constitutes a proof-of-principle study demonstrating that particle physics tests of the ontological status of the quantum state are possible and that $\psi$-epistemic models may exhibit experimentally distinguishable signatures in particle phenomenology.

\end{abstract}



\section{Introduction}
\label{intro}
Since the emergence of quantum mechanics (QM) in the early 20th century, the ontological status of the quantum state has been a subject of deep debate.
Although these discussions and disagreements continue to this day, the emergence of fields such as quantum computing and quantum information has led to the development of new mathematical techniques and experimental methods, enabling the problem to be addressed in a much more thorough and detailed manner. In this context, a well-defined criterion (Harrigan and Spekkens criterion, or for short, the HS criterion) has been established regarding the debate over whether the quantum state corresponds to a physical reality or merely represents the observer’s knowledge about the quantum system \cite{Harrigan,Spekkens}. The HS criterion refers to the former as $\psi$-ontic and the latter as $\psi$-epistemic. Thus, the debate concerning the ontological status of the quantum state has gone beyond being merely philosophical and has become mathematically determinable. Although there are some assertions in the literature that the HS criterion does not provide an appropriate definition \cite{Oldofredi1,Oldofredi2}, the vast majority find the criterion appropriate and use it in their studies \cite{Leifer,Gao1,Gao2}. A highly significant and influential no-go theorem proven through the application of the HS criterion is the Pusey-Barrett-Rudolph (PBR) theorem \cite{Pusey}. The PBR theorem demonstrated that $\psi$-epistemic models have predictions that contradict the predictions of QM.
Therefore, as long as the experimental predictions of QM are correct, $\psi$-epistemic models are excluded. On the other hand, some physicists argue that the epistemic view of the quantum state is necessary to explain a wide range of quantum phenomena, from the collapse of the wave function to the understanding of the huge amount of information hidden in a qubit. They claim that these phenomena provide evidence in favor of the epistemic view (see section 2 of \cite{Leifer} and \cite{Spekkens2}). Therefore, it is important to test the possible predictions of $\psi$-epistemic models that deviate from those of QM. In the literature, various experimental tests have been conducted—and remain an active area of investigation—to examine QM in light of the predictions made by these $\psi$-epistemic models \cite{Patra,Ringbauer,Nigg,Faroughi,Yang}. 

Particle physics is founded on quantum field theory (QFT), and it is evident that a modification of quantum theory could alter all of its consequences. However, uncovering the full implications of a modification originating from QM for particle physics is a very comprehensive and challenging task. On the other hand, in this paper we will demonstrate that a modification of QM based on a $\psi$-epistemic model produces some direct and observable results in polarized particle scattering and decay processes. The implications of $\psi$-epistemic models for particle physics are significant in two respects. First, the large amount of data generated by particle physics experiments can be used to impose constraints on $\psi$-epistemic models through statistical analysis. In other words, particle physics experiments can be employed in the experimental testing of whether the quantum state is ontic or epistemic. Testing a new hypothesis in different physical realms is important to determine its scope and validity. $\psi$-epistemic models have so far been tested using experimental setups based on quantum information processing, but have not been tested using particle physics experiments. Second, numerous studies have been conducted in the literature regarding the existence of new physics beyond the Standard Model (SM) of particle physics, and many alternative models and theories have been proposed. Extensive experimental efforts have been carried out, and continue to be pursued, to test the predictions of these new models and to search for possible signals of new physics beyond the SM \cite{RPP}. Interestingly, a signal that appears to be beyond the SM may not originate from a new particle physics theory or model, but could instead arise from a modification of the nature of QM; more specifically, from a $\psi$-epistemic ontological model. This possibility should not be overlooked, and its implications should be examined.

The aim of the present work is not to derive realistic experimental bounds on $\psi$-epistemic models from particle physics experiments, nor to provide a comprehensive phenomenological analysis under realistic experimental conditions. Rather, our goal is to demonstrate, to our knowledge for the first time, that $\psi$-epistemic models can lead to deviations from standard quantum field-theoretic predictions. In this sense, the present work should be regarded as a proof-of-principle study intended to guide future investigations.
The paper is organized as follows. In the next section, we briefly review the basic formulation of ontological models and give a precise definition of the HS criterion. In Section 3, we discuss the predictions of $\psi$-epistemic models within the context of relativistic QM and derive several generic relations. Section 4 is devoted to the quantum field-theoretic analysis, where the implications of $\psi$-epistemic models for cross sections and decay widths are investigated. The final section presents the conclusions and discussion. After summarizing our results, we discuss the role of relativity in the present work, together with some remarks on relativistic models.

\section{A Brief Overview Of Ontological Models And $\bf{\Psi}$-Ontology \\Theorems}

Ontological models assume that every physical system possesses a real physical state, referred to as an {\it ontic state}. Such an ontic state corresponds to an element of the ontic state space $\Lambda$ and is denoted by $\lambda$ \cite{Harrigan}.\footnote {Here, we will follow the formalizm of Ref.\cite{Harrigan}.} When a quantum system is prepared in a quantum state $|\psi>$, it occupies a specific ontic state $\lambda \in \Lambda$. However, the preparation procedure may not determine the ontic state uniquely. Therefore, an appropriate representation of our knowledge about the ontic state is given by a probability distribution $\mu(\lambda)$ over the space $\Lambda$. To provide a precise and mathematical definition, we consider the $\sigma$-algebra $\Sigma$ of subsets of $\Lambda$ and construct the measurable space $(\Lambda, \Sigma)$. Our knowledge about the preparation $\psi$ is represented by a probability measure $\boldsymbol{\mu_\psi}:\Sigma \rightarrow [0,1]$ on this measurable space. The measure $\boldsymbol{\mu_\psi}$ can be written in the form $\boldsymbol{\mu_\psi}(\Omega)=\int_\Omega \mu_\psi(\lambda) dm(\lambda)$. Here, $\Omega \in \Sigma$, $\mu_\psi(\lambda)$ is a probability distribution and $m$ is a measure that dominates $\boldsymbol{\mu_\psi}$ \cite{Leifer}. In this paper, we abbreviate $dm(\lambda)$ as $d\lambda$. The support of $\mu_\psi(\lambda)$ is the complement of the set on which it vanishes and is denoted by $Supp(\mu_\psi)$. If we consider an ensemble of systems all prepared in the same quantum state $|\psi>$, then each system in the ensemble is associated with an ontic state lying within the $Supp(\mu_\psi)$. Different members of the ensemble may occupy different ontic states; however, some ontic states may be more likely than others, with their likelihoods determined by the distribution $\mu_\psi(\lambda)$. Let $M$ represent a measurement corresponding to a POVM, and let $E_k \in M$ be one of its elements associated with a measurement outcome indexed by the integer $k$.
For a given measurement $M$ and a preparation $\psi$, the probability of obtaining an outcome that corresponds to $E_k$ is given by \cite{Leifer,Pusey}
\begin{eqnarray}
 \label{prob-onticmodel}
 \text{P}(E_k | M, \psi)=\int_{\Lambda} \xi_{k,M}(\lambda)\mu_\psi(\lambda)d\lambda.
\end{eqnarray}
Here, $\xi_{k,M}(\lambda): \Lambda \rightarrow [0,1]$ is the response function for the measurement satisfying the condition below, which ensures proper normalization.
\begin{eqnarray}
 \label{normalization}
 \sum_{k=1}^d \xi_{k,M}(\lambda)=1 ;\;\;\;\; \forall \lambda
\end{eqnarray}
In the above expression, $d$ is the number of POVM elements. Therefore, for orthonormal measurements, $d$ is equal to the dimension of the Hilbert space. If it is assumed that the ontological model reproduces the predictions of QM, then
\begin{eqnarray}
 \label{QM-onticmodel}
 <\psi|E_k|\psi>=\int_{\Lambda} \xi_{k,M}(\lambda)\mu_\psi(\lambda)d\lambda
\end{eqnarray}
holds for a pure state $|\psi>$.

\begin{figure}[h]
\centering
\includegraphics[scale=1.2]{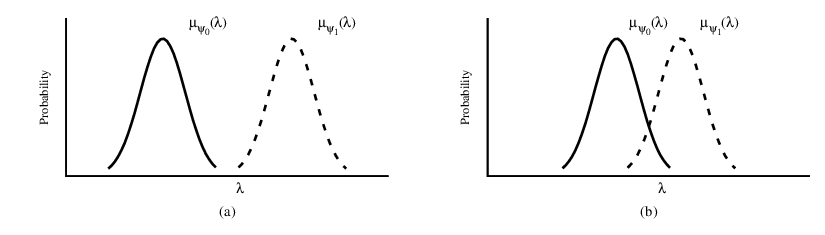}
\caption{Schematic illustration of $\psi$-ontic and $\psi$-epistemic models according to the HS criterion. The distributions $\mu_{\psi_0}(\lambda)$ and $\mu_{\psi_1}(\lambda)$ are non-overlapping in panel (a), which corresponds to a $\psi$-ontic model, and overlapping in panel (b), which corresponds to a $\psi$-epistemic model.\label{fig1}}
\end{figure}

It is possible to classify ontological models into two categories, referred to as $\psi$-ontic and $\psi$-epistemic, based on their stance on the ontological status of the quantum state. In $\psi$-ontic models, the quantum state corresponds to a physical reality, whereas in $\psi$-epistemic models the quantum state represents only the observer’s knowledge about the quantum system. There is a criterion known as the HS criterion \cite{Harrigan}, which is mathematically well-defined and widely accepted in the literature, for determining the $\psi$-ontic/$\psi$-epistemic classification of ontological models. According to the HS criterion, an ontological model is $\psi$-ontic if, for every pair of distinct pure quantum states $|\psi_0>$ and $|\psi_1>$, the corresponding probability distributions $\mu_{\psi_0}(\lambda)$ and $\mu_{\psi_1}(\lambda)$ have disjoint supports, i.e. intersection of their supports is zero measure, $m(Supp(\mu_{\psi_0})\bigcap Supp(\mu_{\psi_1}))=0$  (see Figure \ref{fig1}-(a)). Conversely, an ontological model is $\psi$-epistemic if it is not $\psi$-ontic; that is, there exist two distinct pure quantum states $|\psi_0>$ and $|\psi_1>$ such that the intersection of the supports of the corresponding distributions $\mu_{\psi_0}(\lambda)$ and $\mu_{\psi_1}(\lambda)$ has non-zero measure, $m(Supp(\mu_{\psi_0})\bigcap Supp(\mu_{\psi_1}))>0$ (see Figure \ref{fig1}-(b)). 

In the assessment of a given model under the HS criterion, the {\it total variation distance} and the {\it overlap} play a particularly useful role. The total variation distance between the distributions $\mu_{\psi_0}(\lambda)$ and $\mu_{\psi_1}(\lambda)$ is defined as 
\begin{eqnarray}
 \label{variation distance}
 D(\mu_{\psi_0},\mu_{\psi_1})=\frac{1}{2}\int_{\Lambda} |\mu_{\psi_0}(\lambda)-\mu_{\psi_1}(\lambda)|d\lambda.
\end{eqnarray}
Intuitively, $D(\mu_{\psi_0},\mu_{\psi_1})$ is a measure of how easy it is to distinguish two probability distributions \cite{Pusey}. The overlap for the above probability distributions is given as follows:
\begin{eqnarray}
 \label{overlap}
 w(\mu_{\psi_0},\mu_{\psi_1})=\int_{\Lambda} min \{\mu_{\psi_0}(\lambda),\mu_{\psi_1}(\lambda)\}d\lambda.
\end{eqnarray}
These two important quantities are related by $D(\mu_{\psi_0},\mu_{\psi_1})=1-w(\mu_{\psi_0},\mu_{\psi_1})$; consequently, a model is $\psi$-epistemic if and only if $D(\mu_{\psi_0},\mu_{\psi_1})<1$ or $w(\mu_{\psi_0},\mu_{\psi_1})>0$, $\exists \;\mu_{\psi_0},\mu_{\psi_1}$.

In the literature, a number of no-go theorems prove the $\psi$-ontic nature of ontological models by showing that $\psi$-epistemic models lead to predictions that conflict with those of QM; some of these references are: \cite{Pusey,Hardy,PatraPRL,Aaronson}. These theorems further imply that for any $\psi$-epistemic model to be valid, it would be necessary to deviate from the predictions of QM, and thus to introduce a modification of QM. Each of these theorems is based on its own set of initial assumptions, and the resulting conflict with QM arises under various measurement conditions. In this context, it is instructive to summarize the main argument of the PBR theorem, as it represents both the first and a particularly significant $\psi$-ontology theorem. The PBR theorem has two fundamental premises. The first is the assumption that every quantum system has a real physical state (ontic state), which is a necessary assumption for all ontological models. The second premise is known as the preparation independence postulate, which essentially posits that composite systems prepared in a product state are independent of one another \cite{Leifer,Pusey}. The original proof of the PBR theorem applies to pure quantum states of the following form, with $0<\alpha<\pi/2$, which span two-dimensional Hilbert space:
\begin{eqnarray}
 \label{quantumstates1}
 |\psi_0>=\cos{\frac{\alpha}{2}}\;|0>+\sin{\frac{\alpha}{2}}\;|1>\\
 \label{quantumstates2}
 |\psi_1>=\cos{\frac{\alpha}{2}}\;|0>-\sin{\frac{\alpha}{2}}\;|1>.
 \end{eqnarray}
In the PBR theorem, the assumption of $\psi$-epistemicity is adopted as a reductio ad absurdum hypothesis, and it is shown that this assumption leads to results that are incompatible with the predictions of QM. The states given in (\ref{quantumstates1}) and (\ref{quantumstates2}) are not orthogonal, and therefore are not perfectly distinguishable by measurement. On the other hand, the following states, constructed from n independent preparations,
\begin{eqnarray}
 \label{compositestates1}
|\Psi_{\vec x}>=|\psi_{x_1}>\otimes|\psi_{x_2}>\otimes...\otimes|\psi_{x_n}>
\end{eqnarray}
are post-Peierls incompatible \cite{Caves}, or, in other terminology, antidistinguishable \cite{Leifer}. Here, $x_j \in \{0,1\}$ and $\vec x=(x_1,x_2,...,x_n)$. Accordingly, by performing a joint measurement on
n prepared systems, it is possible to determine which of the $2^n$ quantum states the composite system is not in. Indeed, the PBR theorem demonstrates that there exist mutually orthogonal entangled measurements such that each measurement yields zero probability for one of the states in (\ref{compositestates1}). More precisely, the entangled measurements introduced in the PBR theorem can be represented by the following orthonormal basis
\begin{eqnarray}
 \label{PBRmeasurement}
|\varsigma_{\vec x}>=(H^{\otimes n}R_\alpha Z_\beta^{\otimes n})^\dagger |x_1,x_2,...,x_n>,\;\;\;<\varsigma_{\vec x}|\varsigma_{\vec x'}>=\delta_{\vec x,\vec x'},\;\;\;\sum_{\vec x}|\varsigma_{\vec x}><\varsigma_{\vec x}|=I
\end{eqnarray}
satisfying the condition
\begin{eqnarray}
 \label{PBRmeasurementzero}
<\varsigma_{\vec x}|\Psi_{\vec x}>=0,\;\;\; \forall \vec x.
\end{eqnarray}
See Ref.\cite{Pusey} for the definitions of the unitary transformations, $H$, $R_\alpha$ and $Z_\beta$. The equations in (\ref{PBRmeasurementzero}) show that, according to QM, the $|\varsigma_{\vec x}>$-measurements on $|\Psi_{\vec x}>$ have zero probability. In the PBR proof, it is deduced that the QM zero-probability prediction expressed in (\ref{PBRmeasurementzero}) necessarily imposes the condition $w(\mu_{\psi_0},\mu_{\psi_1})=0$ on the overlap. (While $w(\mu_{\psi_0},\mu_{\psi_1})=0$ holds in an ideal error-free experiment, under realistic experimental conditions with small amounts of noise one obtains $w(\mu_{\psi_0},\mu_{\psi_1})/2 \leq \epsilon^{1/n}$.) Therefore, $\psi$-epistemic models are excluded. For the purposes of this paper, the contrapositive of this argument is of primary relevance: if a $\psi$-epistemic model correctly describes quantum systems in nature, then there exists at least one measurement $|\varsigma_{\vec {x^\star}}>$ on $|\Psi_{\vec {x^\star}}>$ for which the QM zero-probability prediction fails. Formally, this implies
\begin{eqnarray}
 \label{prob-onticmodel2}
 \text{P}(E_{\varsigma_{\vec x^\star}} | M, \Psi_{\vec {x^\star}})=\int_{\Lambda^n} \xi_{\varsigma_{\vec x^\star},M}(\vec \lambda)\;\mu_{\Psi_{\vec {x^\star}}}(\vec \lambda)d\vec \lambda>0.
\end{eqnarray}
The measurements (\ref{PBRmeasurement}) used in the PBR proof are entangled and act on the composite systems defined in (\ref{compositestates1}). Such measurements typically require complex experimental setups, likely involving the use of quantum circuits. Furthermore, these measurements cannot be performed on a single copy of the system in question. In contrast, particle physics experiments are carried out on ensembles of systems prepared in a specific quantum state, with measurements performed on the individual systems within each ensemble. Therefore, it does not appear feasible to draw conclusions for particle physics from the PBR theorem. However, there are various $\psi$-ontology theorems that can be used for this purpose. Theorems of this kind rely on distinct initial assumptions and, as a result, are applicable to different classes of $\psi$-epistemic models. Let us summarize some of them. Hardy established a no-go theorem for $\psi$-epistemic models fulfilling the condition referred to as {\it ontic indifference} \cite{Hardy}. According to the ontic indifference condition, any quantum transformation on a system that leaves a given pure state $|\psi>$ invariant does not change the ontic states ($\lambda$) within the support of the corresponding distribution. As shown in the Supplementary Material of Ref.\cite{PatraPRL}, the ontic indifference condition for $\psi$-epistemic models requires the existence of shared ontic states for orthogonal quantum states. Accordingly, the intersection of the supports of the probability distributions $\mu_{\psi}(\lambda)$ and $\mu_{\psi'}(\lambda)$ corresponding to states satisfying $<\psi|\psi'>=0$ has a non-zero measure. This clearly leads to a deviation from the zero-probability predictions of QM. In $\psi$-epistemic models that satisfy this condition, such deviations from the QM zero-probability predictions can arise in measurements performed on a single quantum system, and the result holds for Hilbert space dimensions $d\geq2$. In fact, the ontic indifference condition is a rather strong condition. For every orthogonal basis of the Hilbert space, the supports of the corresponding distributions intersect on a set of non-zero measure. Accordingly, given any orthonormal measurement basis $\mathcal{B}=\{|i>\}_{i=1}^d$ in a
$d$-dimensional Hilbert space, there exists at least one measurement $|j>$ for which, when $|j>$ is considered as the measurement and the remaining
$d-1$ orthogonal basis vectors as preparations, a deviation from the zero-probability prediction of QM can occur. Hardy also examines models that satisfy a condition called {\it restricted ontic indifference} \cite{Hardy}. Within this class of $\psi$-epistemic models as well, a deviation from the QM zero-probability prediction occurs. However, for models in this class, the contradiction with the QM prediction arises in a Hilbert space extended by an ancilla \cite{Hardy,PatraPRL}. Therefore, a deviation from the QM zero-probability prediction has not been shown in a measurement performed on a single quantum system with Hilbert space dimension $d\geq2$.\footnote{It should be noted that it cannot be concluded for $\psi$-epistemic models satisfying restricted ontic indifference that measurements on single quantum systems never deviate from the QM zero-probability prediction; rather, there is simply no proof establishing that such a deviation occurs. See Ref.\cite{PatraPRL} Supplementary Material section 2 proof of theorem 2.}

Another important class of $\psi$-ontology theorems, established by Patra et al.\cite{PatraPRL}, addresses {\it $\delta$-continuous} and {\it continuous} $\psi$-epistemic models. Let $B(\delta,\psi)=\{|\phi> :\left|<\phi|\psi>\right|\geq1-\delta \}$ denote a $\psi$-centered ball of radius $\delta>0$. A model is said to be $\delta$-continuous if, for any preparation $\psi$ there exists an ontic state $\lambda$ such that, for every preparation $|\phi'>\in B(\delta,\psi)$, we have $\mu_{\phi'}(\lambda)>0$. A model is continuous if it is $\delta$-continuous for some $\delta>0$. $\delta$-continuous $\psi$-epistemic models with $\delta \geq 1-\sqrt{\frac{d-1}{d}}$ contradict QM, as they predict deviations from the QM zero-probability predictions. Here, $d$ denotes the dimension of the Hilbert space.  In $\delta$-continuous models deviations from the QM zero-probability predictions can arise in measurements performed on a single quantum system. To be precise, for any orthonormal basis $\mathcal{B}=\{|j>\}_{j=1}^d$, there exist states $|\psi_j>$ satisfying $<j|\psi_j>=0$ $\forall j$, such that, for at least one $k$, the projective measurement onto $|k>$, when applied to $|\psi_k>$, deviates from the zero-probability prediction of QM, i.e. $\text{P}(E_k | M, \psi_k)>0$, $\exists k$ (See Theorem 1 of \cite{PatraPRL}). Here, $|\psi_k>$ is not an element of the basis $\mathcal{B}=\{|j>\}_{j=1}^d$. As the dimension of the Hilbert space increases, the constraint imposed by the $\delta$-continuity condition on $\psi$-epistemic models becomes weaker. Indeed, for $d=2$, the radius of the $\psi$-centered ball is $\delta=1/\sqrt {2}$, while in the limit $d\to \infty$, the radius approaches zero. Therefore, as the dimension of the Hilbert space of the preparations increases, the overlap of the corresponding distributions tends to decrease. We note that by a 'weaker condition' we mean the enlargement of the class of $\psi$-epistemic models satisfying this condition; that is, as the condition weakens, the number of models within the class increases. Therefore, if one aims to rule out $\psi$-epistemic models under the assumption that the predictions of QM are correct, a weaker condition corresponds to a stronger exclusion. In the case of continuous models, it has been shown that, when combined with an assumption known as separability, they contradict the predictions of QM. As before, this contradiction manifests itself as deviations from zero-probability predictions. However, deviations from the predictions of QM appear in measurements on tensor-product states. In contrast to the tensor product in Eq. (\ref{compositestates1}) used in the proof of the PBR theorem, the situation considered here involves the tensor product of identical copies of the same system, denoted (following the notation of Ref.\cite{PatraPRL}) as $|\phi_k^{\otimes n}\rangle$, where entangled measurements may still be present. (For details, see the proof of Theorem 2 in Ref.\cite{PatraPRL}.) Consequently, a deviation from the QM prediction is not guaranteed in measurements on an ensemble of identically prepared $|\phi_k>$ states.

Finally, we summarize the {\it symmetric} and {\it strongly symmetric, maximally nontrivial} $\psi$-epistemic models, which are relevant for the purposes of this paper. Maximally nontrivial models are those in which the probability distributions corresponding to all non-orthogonal quantum states have non-zero overlaps. In other words, if $|\psi_0\rangle$ and $|\psi_1\rangle$ are not orthogonal, then $w(\mu_{\psi_0},\mu_{\psi_1})>0$. A model is symmetric if it satisfies the following two conditions \cite{Aaronson}:
\begin{enumerate}
    \item the ontic state space is the complex projective space, $\Lambda = \mathbb{CP}^{d-1}$,
    \item any unitary transformation that leaves a quantum state $|\psi\rangle$ invariant leaves the corresponding distribution $\mu_\psi$ invariant, $U|\psi\rangle=|\psi\rangle\;\Rightarrow \mu_{U\psi}(U\lambda)=\mu_{\psi}(\lambda)$.
\end{enumerate}
Condition 1. implies that ontic states are associated with quantum states defined on the projective Hilbert space. This condition further implies that the ontic state space can be endowed with a geometric structure, in which the {\it distance} between two states is characterized by the Fubini–Study metric, $d_{FS}(\psi,\psi')=\frac{2}{\pi} \arccos(|\langle\psi|\psi'\rangle|)$. Turning to the second condition, it is worth noting that it evokes the ontic indifference condition. Upon closer inspection, however, it is evident that this is a weaker requirement than ontic indifference; while ontic indifference ensures that condition 2 is satisfied, the converse does not hold. If condition 2 is imposed not merely for unitary transformations that leave the state $|\psi\rangle$ invariant, but for all unitary transformations, $\mu_{U\psi}(U\lambda)=\mu_{\psi}(\lambda)$, this leads to the definition of strongly symmetric theories. Ref.\cite{Aaronson} demonstrates that symmetric and strongly symmetric maximally nontrivial models can yield outcomes that contradict QM for measurements on a single system with a Hilbert space of dimension $d\geq3$. It has been shown that, for a given measurement basis $M=\{|\phi_i\rangle\}_i^d$ there exists at least one quantum state $|u_j \rangle$ orthogonal to $|\phi_j\rangle$ for which $\int_{B} \xi_{\phi_{j},M}(\lambda)\;\mu_{u_j}(\lambda) d\lambda>0$. (See the proof of Theorem 2 in Ref.\cite{Aaronson}.) Here, $B\subset\Lambda$ denotes a set of non-zero measure that includes a subset of $Supp(\mu_{u_j})$. Consequently, the model yields a prediction that contradicts the zero-probability prediction of QM.

As previously noted, particle physics experiments are conducted on ensembles prepared in a particular quantum state, with measurements performed on individual systems within each ensemble. Therefore, in such experiments, it is not possible to perform entangled measurements on composite systems in the tensor product structure used in some $\psi$-ontology proofs \cite{Pusey}. On the other hand, as summarized above, various $\psi$-epistemic models can
yield predictions that differ from those of the QM at the level of individual measurements performed on the constituent systems of a given ensemble. Indeed, models that satisfy the ontic indifference condition, $\delta$-continuous models, and symmetric or strongly symmetric maximally nontrivial models exhibit this property and produce predictions that can be tested by particle physics experiments. We should also note that when it comes to theorems showing that $\psi$-epistemic models and QM predictions contradict each other through tensor product structures, it cannot be concluded that these theorems imply consistency between the $\psi$-epistemic model and QM in single-system measurements. One may only conclude that these theorems do not furnish such a proof; nevertheless, the lack of such a proof cannot be taken as evidence for its nonexistence. Finally, almost all $\psi$-ontology theorems -including even those whose proofs rely on the tensor product structure- reach a contradiction by demonstrating that $\psi$-epistemic models lead to deviations from the zero-probability predictions of QM. This fact may provide a useful signal for distinguishing $\psi$-epistemic models from various models and theories beyond the SM in particle physics experiments.

\section{Some Generic Relations And Relativistic Quantum Mechanics}

Let us now examine more closely the case in which the ontological model is $\psi$-epistemic. To this end, suppose that the supports of the distributions $\mu_{\psi_0}(\lambda)$ and $\mu_{\psi_1}(\lambda)$, corresponding to two distinct pure quantum states $|\psi_0\rangle$ and $|\psi_1\rangle$, have non-zero overlap. Consider a preparation device that prepares a quantum system in the quantum state $\lvert \psi_0 \rangle$. We assume that $n$ such systems are prepared independently of one another. Within the framework of ontological models, the $n$ prepared systems may possess different ontic states. Accordingly, if the ontic state of the first system is denoted by $\lambda_1$, that of the second system by $\lambda_2$, and so on, up to the $n$-th system with ontic state $\lambda_n$, then the states $\lambda_1, \lambda_2, \ldots, \lambda_n$ are distributed over the support of the distribution $\mu_{\psi_0}(\lambda)$. Under the assumption that the distributions have a non-zero overlap, there exists a probability $q>0$ such that the system can be described by both $\mu_{\psi_0}(\lambda)$ and $\mu_{\psi_1}(\lambda)$. This implies that, among the $n$ independently prepared systems, there are $qn$ number of systems whose ontic states are compatible with each of the quantum states $\lvert \psi_0 \rangle$ and $\lvert \psi_1 \rangle$. Indeed, if the prepared systems are measured with a measurement device, the device performs the measurement by accessing the ontic states of the systems. As a result, the device is unable to resolve the quantum state of $q n$ systems and may conclude that they were prepared in the quantum state $\lvert \psi_1 \rangle$. Therefore, if a measurement is performed on the prepared systems in a basis orthogonal to $\lvert \psi_0 \rangle$ (but not orthogonal to $\lvert \psi_1 \rangle$), there is a probability that the measurement device outputs a result because it may identify the system as compatible with $\lvert \psi_1 \rangle$. This situation contradicts the zero-probability prediction of QM.

The above account is heuristic rather than mathematically rigorous. On the other hand, as discussed in the previous section, rigorous no-go results establish that, for certain classes of $\psi$-epistemic models, deviations from the zero-probability predictions of QM arise already in single-system measurements performed on the constituent systems of a given ensemble. It has been proven for the classes of models considered above that the intersection of the supports of at least two distinct distributions corresponding to two distinct quantum states, contains a subset of non-zero measure, say $B$, such that, for some $\lambda \in B$, the response function produces a result that contradicts the zero-probability prediction of QM. Here, the QM zero-probability measurement is orthogonal to one of the quantum states whose support lies within the set $B$. Without loss of generality, let us take two of these states appearing in the proofs as $\lvert \psi_0 \rangle$ and $\lvert \psi_1 \rangle$, and take the zero-probability measurement to be orthogonal to $\lvert \psi_0 \rangle$, denoted by $E_{0\perp}$. Then, we get
\begin{eqnarray}
\label{devzero}
 \xi_{E_{0\perp},M}(\lambda)>0,\; \exists \lambda \in B\; \Rightarrow \text{P}(E_{0\perp} | M, \psi_0)\geq \int_{B} \xi_{E_{0\perp},M}(\lambda)\mu_{\psi_0}(\lambda)d\lambda>0.
\end{eqnarray}
which confirms our heuristic argument. Moreover, it follows from Eq. (\ref{normalization}) that if any of the other measurements exhibits certainty, then a deviation from the QM unit-probability predictions must also occur. For instance, for qubits ($d=2$), any deviation from a zero probability prediction implies a corresponding deviation from certainty;
\begin{eqnarray}
\label{devcertainty}
\xi_{E_{0},M}(\lambda)=1-\xi_{E_{0\perp},M}(\lambda)<1,\; \exists \lambda \in B
\; \Rightarrow \text{P}(E_{0} | M, \psi_0)=\int_{\Lambda} \xi_{E_{0},M}(\lambda)\mu_{\psi_0}(\lambda)d\lambda<1
\end{eqnarray}
where $E_0$ denotes the projector onto $\lvert \psi_0 \rangle$. Let us now consider the measurements $E_k$ performed on the $n$ preparations of the state $\lvert\psi_0\rangle$ discussed in the previous paragraph. According to QM,
\begin{eqnarray}
 n_Q = n \langle\psi_0\lvert E_k \rvert\psi_0\rangle
\end{eqnarray}
of the $n$ measured systems will yield a response to the measurement, whereas in a $\psi$-epistemic model,
\begin{eqnarray}
 n_{\psi} = n \int_{\Lambda} \xi_{E_k,M}(\lambda)\, \mu_{\psi_0}(\lambda)\, d\lambda
\end{eqnarray}
systems will respond to the measurement. The difference between the numbers $n_Q$ and $n_\psi$ can be statistically analyzed. Indeed, it is possible to take the $\psi$-epistemic model as a hypothesis to be tested. The $\psi$-epistemic model can be tested using the $\chi^2$ test, a method widely used in particle physics experiments for hypothesis testing. The $\chi^2$ function is defined as follows \cite{RPP}
\begin{eqnarray}
 \chi^2=\left(\frac{n_Q-n_\psi}{n_Q \delta_{exp}}\right)^2
\end{eqnarray}
where $\delta_{exp}$ represents experimental errors. $\delta_{exp}$ is related to statistical ($\delta_{stat}$) and systematic errors ($\delta_{syst}$) via the  formula, $\delta_{exp}=\sqrt{\delta_{stat}^2+\delta_{syst}^2}$. The statistical error is defined as $\delta_{stat}=\frac{1}{\sqrt {n_Q}}$. Systematic errors, on the other hand, are specific to the experiments performed, and it is not possible to assign a definitive value without reference to a particular experiment. In this study, it is assumed that systematic errors may be reduced to be comparable in magnitude to the statistical errors, and therefore we set $\delta_{syst}=\delta_{stat}$. In the $\chi^2$ hypothesis testing method, the constraint on the tested model at the 95\% confidence level (C.L.) is obtained by imposing the condition
$\chi^2\leq 3.84$ (See \cite{RPP} Table 40.2). Accordingly, the 95\% C.L. bound on the model is given by 
\begin{eqnarray}
\label{Chi2-1}
 |\Delta^{(k)}_{Q\psi}| \leq \left(\frac{7.68 \langle\psi_0\lvert E_k \rvert\psi_0\rangle}{n}\right)^{1/2}
\end{eqnarray}
where, $\Delta^{(k)}_{Q\psi}$ represents the deviation of the measurement outcomes for $E_k$ from the predictions of QM, i.e. $\Delta^{(k)}_{Q\psi}=\langle\psi_0\lvert E_k \rvert\psi_0\rangle-\int_{\Lambda} \xi_{E_k,M}(\lambda)\, \mu_{\psi_0}(\lambda)\, d\lambda$. For $d=2$, in the presence of a deviation from the certainty specified in equation (\ref{devcertainty}), the following bound is obtained:
\begin{eqnarray}
\label{Chi2-2}
\int_{\Lambda} \xi_{E_{0\perp},M}(\lambda)\, min \{\mu_{\psi_0}(\lambda),\mu_{\psi_1}(\lambda)\} d\lambda  \leq \left(\frac{5.42}{n}\right)^{1/2}.
\end{eqnarray}
Here, by making use of the relation $\xi_{E_{0},M}(\lambda)=1-\xi_{E_{0\perp},M}(\lambda)$, the bound has been transferred to the measurement $E_{0\perp}$. It should be noted that, in deriving the constraint given above, the condition
$\chi^2\leq 2.71$ has been adopted. The reason for this choice is that the quantity $\Delta^{(0)}_{Q\psi}$, which represents the deviation from certainty, always takes positive values. Therefore, a one-sided limit can be imposed on
$\Delta^{(0)}_{Q\psi}$, and the corresponding one-sided 95\% C.L. limit for the $\chi^2$ function is set to $2.71$. For the applicability of the $\chi^2$ test, the condition $n_Q \gg 1$ must be satisfied. Therefore, the $\chi^2$ test provides an appropriate method for testing deviations from certainty or from sufficiently large probabilities, provided that the number of preparations $n$ is large. However, the $\chi^2$ method cannot be applied to test deviations from the zero-probability prediction of QM. In the case of $n_Q = 0$, the hypothesis testing method is modified and the analysis is performed under the assumption that the data follow a Poisson distribution. Within this Poisson framework, the limit is directly imposed on $n_\psi$, yielding a 95\% C.L. constraint of $n_\psi \leq 3$ (See \cite{RPP} Table 40.3). Therefore we obtain the following bound:
\begin{eqnarray}
\label{Poisson}
\int_{\Lambda} \xi_{E_{0\perp},M}(\lambda)\, min \{\mu_{\psi_0}(\lambda),\mu_{\psi_1}(\lambda)\} d\lambda  \leq \frac{3}{n}.
\end{eqnarray}
When the constraints in Eqs. (\ref{Chi2-1}) and (\ref{Chi2-2}) are compared with that in Eq. (\ref{Poisson}), it is evident that the constraint in Eq. (\ref{Poisson}) becomes stronger for large values of $n$. On the other hand, it should be borne in mind that these constraints correspond to different measurements. As can be seen from Eqs. (\ref{Chi2-2}–\ref{Poisson}), the limit is imposed not on the overlap itself, but on the integral of the response function multiplied by the distributions’ minimum. The reason is that Eq. (\ref{Chi2-2}) and Eq. (\ref{Poisson}) each involve a single measurement, namely $E_0$ for Eq. (\ref{Chi2-2}) and $E_{0\perp}$ for Eq. (\ref{Poisson}), applied across an ensemble of systems. By combining the constraints corresponding to all measurements in a given measurement basis, it is possible to impose a bound on the overlap using Eq. (\ref{normalization}).

The statistical analysis outlined above is readily applicable to non-relativistic QM experiments. Indeed, non-relativistic QM has been considered during the calculation of QM predictions. However, particle physics applications require quantum states and measurements to be considered within the framework of relativistic QM. Thus, for particle physics experiments, $\chi^2$ and Poisson analyses should employ quantum mechanical predictions provided by relativistic QM or QFT. The ontological model formalism, however, is likewise not a covariant formulation, but rather a non-relativistic one. Does this not then call for a Lorentz-covariant formulation of ontological models? Interestingly, a fully Lorentz-covariant formulation of ontological models is not required for the present analysis. Indeed, as we will show, it is possible to construct the measurement operators as operators acting on non-relativistic QM states. Nevertheless, their physical implementation involves relativistic processes. Therefore, an additional assumption is required, namely that measurements implemented through relativistic processes can still be represented within the ontological framework by well-defined response functions and probabilities. This assumption imposes a much weaker condition than requiring the model to be Lorentz-covariant. Under the above assumption, $\psi$-epistemic models can be treated as experimentally testable hypotheses in particle physics experiments. Any experimentally observable discrepancy between the probability predictions of QM and a $\psi$-epistemic model must persist in all inertial frames of reference. This conclusion does not rely on the Lorentz symmetry of the ontological model itself. The key observation here is that, in the case of measurements performed on an ensemble of particles, the probability predictions of QM or an ontological model are inferred statistically from the frequencies of signals registered by the detectors. Detector signal registrations define physical spacetime events; therefore, any discrepancy between the QM and $\psi$-epistemic model probability predictions in a given frame leads to a similar discrepancy in all frames related by Lorentz transformations.

The spin states of spin-$1/2$ fermions provide a concrete realization of qubit states. We begin by considering non-relativistic spinors and subsequently extend our analysis to relativistic Dirac spinors. In close analogy with the PBR construction, let us choose the spin states $|\psi_0>$ and
$|\psi_1>$, with overlapping support, as in Eqs.(\ref{quantumstates1}) and (\ref{quantumstates2}). Let $\hat{\mathbf{n}} = \cos\alpha\, \hat{\mathbf{z}} + \sin\alpha\, \hat{\mathbf{x}}$, and define the Hermitian operator $\sigma_n = \hat{\mathbf{n}} \cdot \vec{\sigma}$, which is proportional to the component of the spin operator along $\hat{\mathbf{n}}$. It can easily be deduced that
$\sigma_n |\psi_0\rangle = +1\,|\psi_0\rangle \; \text{and}\;
\sigma_n |\psi_{0\perp}\rangle = -1\,|\psi_{0\perp}\rangle$, where $|\psi_{0\perp}>=\sin{\frac{\alpha}{2}}\;|0>-\cos{\frac{\alpha}{2}}\;|1>$ is the state orthogonal to $|\psi_0\rangle$. Therefore, the operator \(\sigma_n\) has the following spectral decomposition $\sigma_n = |\psi_0\rangle \langle \psi_0| - |\psi_{0\perp}\rangle \langle \psi_{0\perp}|$. If a system prepared in the quantum state $|\psi_0\rangle$ exhibits deviations from both a certainty measurement $E_0 \equiv |\psi_0\rangle\langle\psi_0|$ and a zero-probability measurement $E_{0\perp} \equiv |\psi_{0\perp}\rangle\langle\psi_{0\perp}|$, -which necessarily coexist due to the completeness condition $\xi_{E_{0},M}(\lambda)+\xi_{E_{0\perp},M}(\lambda)=1$ (see eqn.(\ref{devcertainty}))-, then there must also be a deviation from the certainty that a measurement of the observable $\sigma_n$ yields the outcome $+1$. {\it That is, when $\sigma_n$ is measured on a system prepared in the state $|\psi_0\rangle$, the outcome $+1$ need not occur with certainty.}

In order to extend our analysis to a relativistic theory, let us consider a Dirac fermion. In the rest frame of the fermion, the unit spinors are eigenvectors of the operator $\vec{\Sigma} \cdot \vec{s}^{\,\prime}$, satisfying
\begin{eqnarray}
\label{RFeigenvalueeqn}
\vec{\Sigma} \cdot \vec{s}^{\,\prime} \, u_{\mathrm{RF}}(\mp s) = \mp \, u_{\mathrm{RF}}(\mp s).
\end{eqnarray}
Here $\vec{\Sigma} = \gamma^5 \gamma^0 \vec{\gamma}$, and in the standard representation it is defined as $\vec{\Sigma} = \begin{pmatrix} \vec{\sigma} & 0 \\ 0 & \vec{\sigma} \end{pmatrix}$. $\vec{\Sigma}$ is commonly referred to as the \(4 \times 4\)  Pauli matrix or the double Pauli matrix. The vector $\vec{s}^{\,\prime}$ defines the spin projection axis in the rest frame of the fermion, and the corresponding spin four-vector is obtained by applying a Lorentz boost to its rest-frame form, $(s^\mu)_{\mathrm{RF}} = (0, \vec{s}^{\,\prime})$, as follows \cite{Itzykson,GreinerQED}:
\begin{eqnarray}
\label{fourspin}
s^\mu=\left(\frac{\vec{p}\cdot\vec{s}^{\,\prime}}{m},\vec{s}^{\,\prime}+\frac{\vec{p}\cdot\vec{s}^{\,\prime}}{m(E+m)}\vec{p}\right).
\end{eqnarray}
Here, $E$, $\vec{p}$, and $m$ denote the energy, three-momentum, and mass of the fermion, respectively. The covariant generalization of eqn (\ref{RFeigenvalueeqn}) is given by the following equation:
\begin{eqnarray}
\label{eigenvalueeqn}
\gamma^5 \, \gamma_\mu \, s^\mu \, u(p, \mp s) = \mp \, u(p, \mp s).
\end{eqnarray}
Using the above equation, the covariant spin projection operator is defined as \cite{Itzykson,GreinerQED}
\begin{eqnarray}
 \label{covariantprojection}
 \hat{P}(s) = \frac{1}{2} ( I + \gamma^5 \gamma_\mu s^\mu ),
\end{eqnarray}
and it can be seen that it satisfies
\begin{eqnarray}
 \label{covarianteigeneqn}
\hat{P}(s)\, u(p, +s) = u(p, +s) \quad \text{and} \quad \hat{P}(s)\, u(p, -s) =0.
\end{eqnarray}
Let us choose the spin projection axis in the rest frame as $\vec{s}^{\,\prime}        =\hat{\mathbf{n}}$, where $\hat{\mathbf{n}}= \cos\alpha\, \hat{\mathbf{z}} + \sin\alpha\, \hat{\mathbf{x}}$ denotes the spin projection axis of the non-relativistic spinors discussed in the previous paragraph. In this case, the rest frame spinors $u_{RF}(\mp s)$ are given in their $4\times 1$ matrix representation by
\begin{eqnarray}
\label{restspinors}
u_{RF}(+s) =\frac{1}{\sqrt 2}\left(
\begin{array}{c}
|\psi_0\rangle \\
|\psi_0\rangle \\
\end{array}
\right), \;\;\;\;\;\;\;\; u_{RF}(-s) =\frac{1}{\sqrt
2}\left(
\begin{array}{c}
|\psi_{0\perp}\rangle \\
|\psi_{0\perp}\rangle \\
\end{array}
\right).
\end{eqnarray}
Here, the spinors are written in the Weyl representation. The relativistic spinors $u(p, \mp s)$ can be obtained by applying Lorentz transformations to the rest frame spinors. For Dirac spinors, Lorentz transformations are implemented using the Lie group $\mathrm{SL}(2,\mathbb{C})$ \cite{Kim}. The $\mathrm{SL}(2,\mathbb{C})$ group has a total of six generators, which we denote by $J_m$ and $K_m$ ($m = 1,2,3$). Here, $J_m$ generate spatial rotations, while $K_m$ generate Lorentz boosts. Without loss of generality, let us assume that the relativistic fermion moves along the $z$-axis in the laboratory frame. Then, the relativistic spinors can be obtained as \cite{Sahin}
\begin{eqnarray}
\label{movingspinors} u(p,\mp s)=&&\exp{(-i\xi
K_3)}\;u_{RF}(\mp s)\nonumber \\=&&\frac{1}{\sqrt
2}\left(\begin{array}{cc}
\cosh{(\frac{\xi}{2})}\mathbb{I}+\sinh{(\frac{\xi}{2})}
\sigma_3 & 0 \\
0 &\cosh{(\frac{\xi}{2})}\mathbb{I}-\sinh{(\frac{\xi}{2})}
\sigma_3 \\
\end{array}
\right)u_{RF}(\mp s)
\end{eqnarray}
where, $\xi$ is the rapidity parameter, defined as $\xi=\frac{1}{2}\ln{\left(\frac{E+|\vec p|}{E-|\vec p|}\right)}$.

An important observation is the following: equation (\ref{RFeigenvalueeqn}) corresponds to the eigenvalue equations of non-relativistic QM when the spinors are expressed in their $4\times 1$ representations. Therefore, projective measurement defined in (\ref{covariantprojection}) reduces, in the rest frame of the fermion, to a measurement on non-relativistic spinors.  If a $\psi$-epistemic model predicts deviations from the definite outcomes $\mp 1$ associated with the observable $\vec{\Sigma}\cdot\vec{s}^{\,\prime}$, the resulting experimentally observable discrepancy between the predictions of QM and the $\psi$-epistemic model persists in all reference frames, as discussed previously. Accordingly, a measurement of the observable $\gamma^5 \gamma_\mu s^\mu$ on a relativistic spinor $u(p, \mp s)$ carries a nonzero probability of deviating from the outcome $\mp 1$. For example, the measurement $\hat{P}(s)$ defined in eqn.(\ref{covariantprojection}) may vanish when acting on $u(p,+s)$, or yield a nonzero result when acting on $u(p,-s)$.

The above discussion was carried out for the realization of qubits in terms of the spin states of spin-1/2 fermions. In the case of qutrits, they can be realized using the spin states of massive spin-1 vector bosons. In the rest frame of the vector boson, let us parametrize three mutually orthogonal spin orientations as follows:
\begin{eqnarray}
\label{spin1rest1}
 \vec{\varepsilon}^{\,\prime}_{1}=&&\left(\cos{\theta^*}\cos{\phi^*},\cos{\theta^*}\sin{\phi^*},-\sin{\theta^*} \right)\\
 \label{spin1rest2}
 \vec{\varepsilon}^{\,\prime}_{2}=&&\left(-\sin{\phi^*},\cos{\phi^*},0 \right)\\
 \label{spin1rest3}
 \vec{\varepsilon}^{\,\prime}_{3}=&&\left(\sin{\theta^*}\cos{\phi^*},\sin{\theta^*}\sin{\phi^*},\cos{\theta^*} \right)
\end{eqnarray}
Here, $\theta^*$ and $\phi^*$ denote the polar and azimuthal angles in the spherical coordinate system, respectively. The spin four-vector in the rest frame is defined as $(\varepsilon^\mu_\lambda)_{\mathrm{RF}} = (0, \vec{\varepsilon}^{\,\prime}_{\lambda})$.
The corresponding spin four-vector in the frame where the vector boson carries momentum $\vec{k}$ is obtained by applying a Lorentz boost to the rest-frame spin vector, in a manner analogous to Eq. (\ref{fourspin}), as
\begin{eqnarray}
\label{fourspinboson}
\varepsilon^\mu_\lambda=\left(\frac{\vec{k}\cdot\vec{\varepsilon}^{\,\prime}_\lambda}{m},\vec{\varepsilon}^{\,\prime}_\lambda+\frac{\vec{k}\cdot\vec{\varepsilon}^{\,\prime}_\lambda}{m(E+m)}\vec{k}\right).
\end{eqnarray}
Equation (\ref{fourspinboson}) represents the polarization of a vector boson with arbitrary spin orientation. The choice $\lambda=1,2,3$ corresponds to the linear polarization basis defined in Eqs. (\ref{spin1rest1})–(\ref{spin1rest3}). On the other hand, the circular polarization basis is defined as \cite{Hagiwara}
\begin{eqnarray}
\label{circularpol}
\varepsilon^\mu_{\pm}=\frac{1}{\sqrt{2}}\left[\mp \varepsilon^\mu_{1}-i\varepsilon^\mu_{2}\right].
\end{eqnarray}
If one chooses $\vec{\varepsilon}^{\,\prime}_{3}=\frac{\vec{k}}{|\vec{k}|}$, the usual helicity basis is obtained. In the helicity basis, the spin four-vector corresponding to $\lambda=3$ describes a longitudinally polarized vector boson with zero helicity. Therefore, instead of $\lambda=3$, we will use the subscript $\lambda=0$ for the spin four-vector associated with longitudinal polarization. The spin four-vectors corresponding to the helicity eigenstates with helicities $+1$ and $-1$ are given in Eq. (\ref{circularpol}), which defines the circular polarization basis, and are denoted by the subscripts $\lambda=\pm$ \cite{Hagiwara}. The longitudinal and transverse polarizations obey the constraints $k \cdot \varepsilon_\lambda=0$ and $\varepsilon^*_{\lambda^\prime} \cdot \varepsilon_{\lambda}=-\delta_{\lambda^\prime\lambda}$; $\lambda^\prime,\lambda=+,-,0$ where, "$\cdot$" represents the Lorentzian inner product with respect to Minkowski metric of trace $-2$ \cite{Ballestrero2018}. Therefore, the projection operator onto a state of definite helicity $\lambda$ is given by $\hat{P}(\lambda)=-\varepsilon^*_{\lambda}$. Its action on a vector field amounts to taking the inner product with the four-vector $-\varepsilon^{*\mu}_{\lambda}$.

A discussion similar to that for spin-1/2 fermions concerning experimentally observable discrepancies between $\psi$-epistemic models and QM can likewise be extended to spin-1 vector bosons. To be precise, for spin measurements, the QM zero-probability prediction can be expressed in a Lorentz-invariant form as $\hat{P}(\lambda) \varepsilon_{\lambda^\prime}=-\varepsilon^*_{\lambda} \cdot \varepsilon_{\lambda^\prime}=0$ for $\lambda\neq\lambda^\prime$. In the rest frame of the vector boson, where a non-relativistic QM description applies, the $\psi$-epistemic model may predict a nonzero probability for this measurement. However, a nonzero probability assigned to an outcome that QM predicts with zero probability gives rise to an experimentally observable discrepancy that persists in all inertial frames.

The analysis presented so far is somewhat incomplete, since it is restricted to relativistic QM and does not incorporate a treatment based on QFT. In the following section, we present a field theoretical analysis of the problem by defining measurements in terms of various particle physics processes.

\section{Implications Of $\bf{\Psi}$-Epistemic Models On Cross sections And \\Decay Widths}

Within the SM of particle physics, the weak interaction is chiral in nature, giving rise to a pronounced correlation between interaction cross sections and particle polarizations. In weak processes, this feature allows particle polarizations to be extracted from cross section measurements, effectively enabling spin measurements of the participating particles. Cross sections are experimentally measurable quantities, as they are directly related to the number of events observed in detectors. Any deviation observed in polarized cross sections may be interpreted as a signal for the existence of a $\psi$-epistemic model, while the absence of such deviations places statistical constraints on the model, thereby excluding the $\psi$-epistemic model. Deviations between $\psi$-epistemic models and the predictions of QM have particularly significant implications in neutrino physics. According to the SM, neutrinos interact with other SM fields only via the weak interaction with minimal coupling. Consequently, neutrino cross sections exhibit a strong dependence on polarization.

In addition to measuring the magnitudes of polarized cross sections, determining the angular distributions of differential cross sections also enables an indirect spin measurement. The chirality dependent nature of the weak interaction implies that spin is not, in general, conserved in weak processes. Nevertheless, conservation of total angular momentum requires the final states to carry orbital angular momentum, giving rise to higher partial wave contributions $\ell \ge1$ in the partial-wave expansion. Consequently, for many weak interaction processes, the angular distribution of the differential cross section encodes information about the particle polarizations. Hence, the particle polarizations can be extracted through measurements of the angular distributions of the differential cross sections. This method is commonly used, especially for measuring the polarization of $W$ and $Z$ bosons \cite{Ballestrero2018,Mirkes1994,L3coll2003,CMScoll2011,Atlascoll2012,CMScoll2015,Atlascoll2016,Atlascoll2023}.

\subsection{Qubit case: spin 1/2 fermions}

The pronounced dependence of neutrino cross sections on polarization renders neutrino physics a powerful probe for particle physics tests of $\psi$-epistemic model predictions in the context of qubit systems. Neutrinos couple to other SM fields via weak interaction through a vector-axial vector ($V-A$) vertex. This interaction projects the neutrino field onto its left-handed component. Consequently, in any particle process in which neutrinos are produced, the resulting neutrinos are left-handed. This fact can be expressed mathematically as $ \hat{L}\, u_\nu(p) = u_\nu(p) $, where $ \hat{L} = \frac{1}{2}(1 - \gamma^5) $ denotes the left-handed projection operator and $ u_\nu(p) $ represents the spinor associated with the neutrino field. It is well established that massless fermions are 100\% longitudinally polarized \cite{Wigner}. For massless fermions, the chirality and helicity states coincide, and left-handed massless fermions produced in any SM process are in a pure negative helicity state. Indeed, for helicity states, the spin four-vector is obtained from Eq. (\ref{fourspin}) by taking $ \vec{s^\prime}=\lambda \frac{\vec{p}}{|\vec{p}|}$, resulting in
\begin{eqnarray}
\label{fourhelicity}
s^\mu_{(\lambda)}=\lambda\left(\frac{|\vec{p}|}{m},\frac{E}{m}\frac{\vec{p}}{|\vec{p}|}\right).
\end{eqnarray}
Here, $\lambda = +1, -1$ defines positive and negative helicity, respectively. In the ultrarelativistic limit, since $s^\mu(\lambda) \to \lambda\; \frac{p^\mu}{m}$, the projection operator acting on the spinor behaves as
\begin{eqnarray}
\label{ultrarellimit}
\hat{P}(s_{(\lambda)}) u(p, \lambda) \to \frac{1}{2}(I+\frac{\lambda}{m} \gamma^5 \gamma_\mu p^\mu) u(p, \lambda) = \frac{1}{2}(I+\lambda \gamma^5)u(p, \lambda).
\end{eqnarray}
The final equality follows from the Dirac equation, and as such, holds exclusively for fields satisfying the on-shell condition. As seen from Eq.(\ref{ultrarellimit}), in the ultrarelativistic limit the spin projection operators for negative and positive helicity coincide with the left- and right-handed chiral projection operators, respectively. Accordingly, one may regard the weak interaction as furnishing a natural preparation mechanism for neutrino states of negative helicity. However, the identification of helicity with chirality is strictly valid only in the ultrarelativistic limit. Experimental evidence from Super-Kamiokande \cite{SuperKamiokande} and the Sudbury Neutrino Observatory \cite{SNO} has established that neutrinos undergo flavor oscillations and therefore have nonzero masses. Although neutrinos are not strictly massless, their masses are exceedingly small. Consequently, the energy scales of many present-day experiments are much higher than the neutrino mass. Hence, it is often an excellent approximation to treat them as massless and purely negative helicity states when computing cross sections, with exceptions in certain low-energy experiments, such as those dedicated to mass measurements or those sensitive to neutrino mass effects. In practice, high energy neutrino production processes, which are relevant to many present-day experiments, are thus taken to furnish an almost perfectly pure preparation of negative helicity neutrino spin states.

Detection of the produced neutrinos constitutes a measurement on their spin states. Since their interaction with the detector constituents is proportional to the left-chiral projection operator $\hat L$, the detection process acts as a projector. Consequently, a neutrino prepared in a negative helicity state is projected onto the eigenstate with the same helicity, which in the ultrarelativistic limit coincides with a left-chiral state. Therefore, the measurement of the neutrino spin yields a {\it certain} outcome according to QM (within the ultrarelativistic approximation). On the other hand, a $\psi$-epistemic model can produce predictions that depart from certainty, as indicated by equation (\ref{devcertainty}). Intuitively, this can be understood through the following observation: Assume that, in the rest frame of the neutrino, the distribution $\mu_{\psi_0}$ associated with the spin state $|\psi_0\rangle$ corresponding to negative helicity has a non-zero overlap with another distribution $\mu_{\psi_1}$ corresponding to a distinct state $|\psi_1\rangle \neq |\psi_0\rangle$. Here, $|\psi_0\rangle$ and $|\psi_1\rangle$ denote non-relativistic $2\times1$ spinors. In the neutrino’s rest frame, these spinors can be embedded into $4\times1$ representations, $u_{RS(0)}$ and $u_{RS(1)}$, as in Eq. (\ref{restspinors}). Due to the non-zero overlap, there exists a positive probability $q > 0$ that the system (the neutrino spin), in its rest frame, is described by ontic states lying in the support of both $\mu_{\psi_{0}}$ and $\mu_{\psi_{1}}$. Since the detector performs the measurement by accessing the ontic state of the neutrino, it cannot discriminate between the corresponding quantum spin states. It may therefore infer that the neutrino spin state in the rest frame is compatible with $u_{RS(1)}$ rather than $u_{RS(0)}$. But the Lorentz-transformed spinor $u_{\nu}(p) = \exp(-i \xi K_3)\, u_{RF(1)}$
is not an eigenstate of $\hat{L}$. Therefore, in the laboratory frame, the detector may generate an output that deviates from a definite outcome for the chirality, i.e. the cross section may be reduced compared to the predictions of QFT. We would like to emphasize that, in the course of the preceding heuristic argument, we did not assume any overlap between the supports of the $\mu_\psi$ distributions associated with relativistic spinors; rather, our analysis was carried out solely within the framework of a non-relativistic ontological model, supplemented by the additional assumption that the measurements involving relativistic processes are well defined.

We now proceed to construct the argument in a more precise and rigorous way. Consider a neutrino beam incident on the detector. These neutrinos may undergo interactions with the neutrons, protons, and electrons in the detector. Without loss of generality, we focus on the neutrino absorption channel in which the incident neutrinos are converted into charged leptons. Such a process involves a charged current of the form
\begin{eqnarray}
\label{chargedcurrent1}
J^\mu=\bar u_\ell \Gamma^\mu \hat L u_\nu(p).
\end{eqnarray}
Without loss of generality, let's assume the neutrino beam moves along the $+z$- axis in the laboratory (lab.) frame. In the rest frame of the neutrino, the spinors for an arbitrary spin orientation with spin projection axis $\vec{s}^{\,\prime}= \cos\alpha\, \hat{\mathbf{z}} + \sin\alpha\, \hat{\mathbf{x}}$ are given by (\ref{restspinors}). Here, the dependence on the azimuthal angle has been omitted due to azimuthal symmetry. Accordingly, the Lorentz-boosted relativistic spinor in the lab frame is given by the relation (\ref{movingspinors}) as
\begin{eqnarray}
\label{generalspinors1} u(p,+s)=\frac{1}{\sqrt 2}\left(
                                     \begin{array}{c}
                                       \cos{\frac{\alpha}{2}}\; e^{\xi/2} \\
                                       \sin{\frac{\alpha}{2}}\; e^{-\xi/2} \\
                                       \cos{\frac{\alpha}{2}}\; e^{-\xi/2} \\
                                       \sin{\frac{\alpha}{2}}\; e^{\xi/2} \\
                                     \end{array}
                                   \right).
                                   \end{eqnarray}
For $u(p,-s)$, substitute $\alpha \to \alpha + \pi$. One may then write the charged current (\ref{chargedcurrent1}) for a neutrino with arbitrary spin orientation as
\begin{eqnarray}
\label{chargedcurrent2}
J^\mu=e^{-\xi/2}\cos{\frac{\alpha}{2}}\; {J_{RF}^{(+)}}^\mu + e^{\xi/2}\sin{\frac{\alpha}{2}}\; {J_{RF}^{(-)}}^\mu
\end{eqnarray}
where,
\begin{eqnarray}
\label{chargedcurrent3}
{J_{RF}^{(\pm)}}^\mu=\bar u_\ell \Gamma^\mu \hat L\; {u_\nu}_{RF}^{(\pm)}.
\end{eqnarray}
In the above expression, ${u_\nu}_{RF}^{(\pm)}$ denote the spinors corresponding to spin up and spin down along the $z$-axis in the neutrino rest frame. Let us note that the angle $\alpha$ in the current (\ref{chargedcurrent2}) specifies the spin orientation with respect to the neutrino rest frame. Therefore, the neutrino interaction provides a measurement of the spin orientation in the neutrino rest frame. Explicitly, $\alpha = 0$ corresponds to positive helicity states, which in the rest frame coincide with spin up along the $z$-axis, while $\alpha = \pi$ corresponds to negative helicity states, which in the rest frame coincide with spin down along the $z$-axis. From (\ref{chargedcurrent2}), it follows that in the lab frame, where $\xi \gg 1$, the current component associated with positive helicity is strongly suppressed, whereas the component associated with negative helicity remains large. Accordingly, the magnitude of the cross section observed in the lab frame effectively distinguishes between spin-up and spin-down states in the rest frame.\footnote{Recall that in the lab frame the $z$-axis is chosen to coincide with the direction of motion of the neutrino beam. Consequently, the cross section measurement effectively measures the spin states in the neutrino rest frame with respect to the detector–neutrino relative velocity direction. Furthermore, it is useful to recall that the total cross section, the number of events observed in the detector, and the chirality operator are Lorentz invariant quantities.} Namely, a large value of the cross section consistent with QFT indicates that the neutrinos are in the negative-helicity state (spin-down in the rest frame), whereas a value too small to be observed indicates that they are in the positive-helicity state (spin-up in the rest frame). From our discussion in Section 2, we know that some $\psi$-epistemic models --for example, those satisfying the ontic indifference condition and the $\delta$-continuous models with $\delta \ge 1-\frac{1}{\sqrt{2}}$-- yield predictions that deviate from the zero-probability predictions of QM already at the level of two-dimensional Hilbert spaces. Such a deviation, by virtue of relation (\ref{devcertainty}), also leads to a deviation from the certainty prediction of QM. Therefore, $\psi$-epistemic models of this type predict a reduction in neutrino cross sections. As we will discuss, the measurement of the cross section admits an interpretation in terms of a POVM element, and a deviation from unit probability corresponds to a suppression of the cross section. Before addressing the details of this issue, we need to clarify an additional point. While the current in (\ref{chargedcurrent1}) is instrumental in the determination of the cross section, it is not itself directly proportional to it. The more fundamental quantity in this context is the scattering amplitude. To this end, let us consider the scattering amplitude for the neutrino absorption process $\nu a \to \ell b$. Here, $a$ and $b$ denote charged fermions (quarks or baryons) and $\ell$ denotes a charged lepton. The tree-level amplitude for this process can be written in the form
\begin{eqnarray}
\label{amplitudechargedcurrent1}
M=g\;J_1^\mu\;J_{2\mu}
\end{eqnarray}
where $J_1^\mu=\bar u_{\ell} \gamma^\mu \hat L u_\nu$ is the charged current that contains the neutrino field and $J_2^\mu$ is the charged current for charged fermions and $g$ is the coupling constant.\footnote{For brevity, we work in the regime $\sqrt{s} \ll m_W$, where the $W$-boson propagator can be approximated as $\frac{g_{\mu\nu} - q_\mu q_\nu/m_W^2}{q^2 - m_W^2} \approx -\,\frac{g_{\mu\nu}}{m_W^2}$. Nevertheless, our main result remains valid independently of this approximation.} By substituting the spinor in (\ref{generalspinors1}), representing a neutrino with an arbitrary spin orientation in its rest frame, into the current $J_1^\mu$, one obtains the following relations:
\begin{eqnarray}
\label{chargedcurrent-spinup} J_1^\mu(+s)&&=\sin{\left(\frac{\alpha}{2}\right)}\left(J_1^\mu\mid_{m_\nu=0}\right)+\mathcal{O}(e^{-\xi})\times\left(J_1^\mu\mid_{m_\nu=0}\right)\\
\label{chargedcurrent-spindown} J_1^\mu(-s)&&=\cos{\left(\frac{\alpha}{2}\right)}\left(J_1^\mu\mid_{m_\nu=0}\right)+\mathcal{O}(e^{-\xi})\times\left(J_1^\mu\mid_{m_\nu=0}\right)
\end{eqnarray}
For massless neutrinos, the current is entirely determined by the left-handed spinor, so that in the relations above $J_1^\mu\mid_{m_\nu=0}=\bar u_{\ell} \gamma^\mu \hat L u^{(L)}_\nu$. Therefore, the squared amplitude takes the form:
\begin{eqnarray}
\label{squaredamplitude1} |M(\pm s)|^2=\frac{\left(1\mp\cos\alpha\right)}{2}\left(|M|^2\mid_{m_\nu=0}\right)+\mathcal{O}(e^{-\xi})\times \left(|M|^2\mid_{m_\nu=0}\right).
\end{eqnarray}
As can be seen from (\ref{squaredamplitude1}), the magnitude of the squared amplitude -and hence of the cross section- depends on the angle $\alpha$, which determines the spin orientation of the neutrino in its rest frame. For $\alpha = 0$ (corresponding to positive helicity in the lab frame), the squared amplitude is strongly suppressed, whereas for $\alpha = \pi$ (corresponding to negative helicity in the lab frame), it attains a relatively large value. Therefore, as discussed previously, the measurement of the cross section effectively constitutes a measurement of the spin orientation in the neutrino rest frame.

In general, polarized cross section measurements cannot be interpreted as projective measurements on initial particle polarization states. In many processes -such as the neutrino absorption process- the initial state particles are destroyed, and thus the initial spin states are not subject to a projective measurement. On the other hand, in interactions involving polarized particles, cross section measurements may be regarded as POVM elements acting on the initial polarisation states. To illustrate this in more detail, let us reconsider the SM process $\nu a \to \ell b$. For this process, we define the following positive operators
\begin{eqnarray}
\label{neutrinoPOVM1}
\mathcal{E}_L=\frac{{M^L_{s_\nu}}^\dagger M^L_{s_\nu}}{\left(|M^L|^2\mid_{m_\nu=0}\right)}\;,\;\;\;\mathcal{E}_R=\frac{{M^R_{s_\nu}}^\dagger M^R_{s_\nu}}{\left(|M^R|^2\mid_{m_\nu=0}\right)}
\end{eqnarray}
where
\begin{eqnarray}
\label{neutrinotransitionop}
M^L_{s_\nu}=g\;J_{2\mu}\left(\bar u_{\ell} \gamma^\mu \hat L \exp(-i \xi K_3)\right)\;,\;\;M^R_{s_\nu}=g\;J_{2\mu}\left(\bar u_{\ell} \gamma^\mu \hat R \exp(-i \xi K_3)\right).
\end{eqnarray}
Here $J_{2\mu}$ denotes the charged current for charged fermions (see (\ref{amplitudechargedcurrent1})), while $|M^L|^2\mid_{m_\nu=0}$ and $|M^R|^2\mid_{m_\nu=0}$ denote the unpolarized squared amplitudes for massless neutrinos. The superscripts $L$ and $R$ are introduced to indicate that the neutrino interaction is proportional to the left-handed and right-handed chiral projections $\hat L$ and $\hat R$, respectively. In the SM, the neutrino coupling is known to be left-handed. Consequently, $\mathcal{E}_R$ does not correspond to a SM interaction; it is defined solely on mathematical grounds to ensure the completeness of the POVM set. Observe that the rest frame spinor ${u_\nu}_{RF}$ associated with the incoming neutrino does not enter the matrix elements (\ref{neutrinotransitionop}). It follows that $M^L_{s_\nu}$ and $M^R_{s_\nu}$ define linear functionals on the neutrino spin Hilbert space $\mathcal{H}_{s_\nu}$,  while $\mathcal{E}_L$ and $\mathcal{E}_R$ are operators on $\mathcal{H}_{s_\nu}$, i.e., $M^L_{s_\nu},M^R_{s_\nu}: \mathcal{H}_{s_\nu}\rightarrow \mathbb{C}$;  $\mathcal{E}_L,\mathcal{E}_R: \mathcal{H}_{s_\nu}\rightarrow \mathcal{H}_{s_\nu}$. (The subscript $s_\nu$ in (\ref{neutrinotransitionop}) does not label a spin component. Rather, it indicates that these linear functionals are defined on the neutrino spin Hilbert space $\mathcal{H}_{s_\nu}$.) The operators $\mathcal{E}_L$ and $\mathcal{E}_R$ are not necessarily diagonal in the incoming neutrino spin basis. By taking the diagonal elements of these operators in the incoming neutrino spin basis $\{|s_\nu \rangle_\text{in}\}_{s_\nu}$, we construct the following operators:
\begin{eqnarray}
\label{neutrinoPOVM11}
E_{L(R)}=\sum_{s_\nu}\; {}_{\text{in}}\langle s_\nu|\mathcal{E}_{L(R)} |s_\nu\rangle _\text{in}\;|s_\nu \rangle_\text{in}\; {}_{\text{in}}\langle s_\nu|
\end{eqnarray}
Proceeding as in the derivation of Eq. (\ref{squaredamplitude1}) and using the spinor (\ref{generalspinors1}), one finds by computing the squared amplitudes for the $L$ and $R$ interactions that
\begin{eqnarray}
\label{neutrinoPOVM2}
E_L=E_{-}+\mathcal{O}(e^{-\xi})\;,\;\;\;E_R=E_{+}+\mathcal{O}(e^{-\xi})
\end{eqnarray}
,where
\begin{eqnarray}
\label{neutrinoPOVM3}
E_{-}=\frac{\left(1-\cos\alpha\right)}{2}\;|s_\nu \rangle \langle s_\nu|+\frac{\left(1+\cos\alpha\right)}{2}\;|-s_\nu \rangle \langle -s_\nu| \\
E_{+}=\frac{\left(1+\cos\alpha\right)}{2}\;|s_\nu \rangle \langle s_\nu|+\frac{\left(1-\cos\alpha\right)}{2}\;|-s_\nu \rangle \langle -s_\nu|.
\end{eqnarray}
Here $|s_\nu \rangle$ and $|-s_\nu \rangle$ denote the spin-up and spin-down neutrino spin states with respect to the spin-projection axis of the incoming neutrinos, whose orientation is determined by the angle $\alpha$ (here and henceforth, the subscript "in" is omitted for brevity). Clearly, $E_{-}$ and $E_{+}$ are positive operators satisfying $E_{-}+E_{+}=I$. Furthermore, for a spin state prepared in $|\pm s_\nu\rangle$, the probabilities that a helicity measurement yields the outcomes $\lambda=\pm1$ are given by
\begin{eqnarray}
\label{neutrinoPOVMProb1}
P(\lambda=\pm1 \mid |s_\nu\rangle)=\langle s_\nu| E_{\pm} |s_\nu\rangle=\frac{\left(1\pm\cos\alpha\right)}{2}\\
\label{neutrinoPOVMProb2}
P(\lambda=\pm1 \mid |-s_\nu\rangle)=\langle -s_\nu| E_{\pm} |-s_\nu\rangle=\frac{\left(1\mp\cos\alpha\right)}{2}.
\end{eqnarray}
It follows that $E_{+}$ and $E_{-}$ form the POVM elements associated with the neutrino helicity measurement. For ultrarelativistic neutrinos, $E_L$ and $E_R$ approach $E_{-}$ and $E_{+}$, respectively, to a very high degree of accuracy and hence effectively constitute POVM elements. The correspondence becomes exact in the limit $\xi \to \infty$. Since neutrino interactions in the SM are purely left-handed, cross section measurements cannot realize a complete measurement (i.e., the full POVM set), but instead implement the measurement determined by $E_L$. This does not imply that an $E_R$ measurement cannot exist; rather, it cannot be realized through cross section measurements in SM interaction processes. As seen from Eqs. (\ref{neutrinoPOVMProb1}) and (\ref{neutrinoPOVMProb2}), an $E_L$ measurement performed on a neutrino beam prepared in the negative helicity state yields the outcome with probability $\approx 1$. Consequently, any $\psi$-epistemic model that violates the certainty prediction of QM yields a reduced polarized cross section for negative-helicity neutrinos in the SM.

Finally, we clarify a subtle point. In the discussion above, the $E_L$ measurement realized via the neutrino cross section is only an approximate POVM element --though it provides a good approximation for many neutrino physics experiments-- and becomes exact only in the limit $\xi\to \infty$. This should not be taken to mean that $E_L$ does not represent a genuine measurement for neutrinos with nonzero, albeit very small, masses. The essential point is that no measurement device is error-free: any realistic measurement necessarily involves some degree of noise. Accordingly, $E_{-}$ can be viewed as the ideal POVM element, whereas $E_L$ corresponds to a noisy realization of it, with the discrepancy between the two lying within experimental uncertainties. Indeed, the difference between the measurements corresponding to $E_{-}$ and $E_L$ can be made parametrically small relative to experimental uncertainties by considering sufficiently high-energy neutrino experiments.

\subsection{Qutrit case: spin 1 vector bosons}
The spin states of massive spin-1 particles span a three-dimensional Hilbert space and provide physical realizations of qutrits. The $W$ and $Z$ bosons, which mediate the weak interaction, constitute examples of massive spin-1 particles. Numerous studies and experiments have been conducted on the experimental determination of their polarization states (see, e.g.,\cite{Ballestrero2018,Mirkes1994,L3coll2003,CMScoll2011,Atlascoll2012,CMScoll2015,Atlascoll2016,Atlascoll2023}), and this topic remains an active area of research. Determining the polarization states of $W$ and $Z$ bosons is carried out through the measurement of the angular distribution of their decay products. Consequently, the angular distribution measurement effectively serves as a measurement of the bosons' polarization states. The correlation between the polarization states of $W$ and $Z$ bosons and the angular distribution of their decay products stems from the fact that total angular momentum is conserved in the decay processes. It is well known that, due to the $V-A$ structure of the weak interaction, spin is not, in general, conserved, i.e., spin projection along a given direction is not conserved. However, total angular momentum must be conserved; accordingly, any difference between the initial and final spin projections along a given direction is compensated by the orbital angular momentum of the final-state particles. This, in turn, affects the angular distribution of the final-state particles through the contribution of partial waves with $\ell \ge 1$.

Let us examine the details within the context of the $W^-\to \ell \bar{\nu}_\ell$ ($\ell=e^-, \mu^-, \tau^-$) decay channel. For this decay channel, the tree-level Feynman amplitude takes the form
\begin{eqnarray}
\label{amplitudeWdecay}
M=\frac{g_W}{\sqrt 2}\bar u(k_1,s_1)\gamma_\mu \hat L v(k_2,s_2)\varepsilon^\mu(k,\lambda_W)
\end{eqnarray}
where $k_1(k_2)$ is the momentum of the charged lepton (anti-neutrino) and $s_1,s_2$ are their respective spin indices. $k$ and $\lambda_W$ denote the momentum and polarisation of the $W$ boson. The helicity amplitudes and the differential decay rate for this process are evaluated in the $W$ boson rest frame and are provided in Appendix A. As discussed, measuring the angular distribution of lepton events from $W$ decay allows for the determination of the $W$ boson polarization. This is evident from Figure \ref{fig2}, where the differential decay rates corresponding to $\lambda_W=+1,-1$, and $0$ have markedly different angular distributions. The observation of signals within specific angular intervals is crucial for isolating particular polarization states. For example, a signal detected in the $\theta\in[0, \epsilon]$ range corresponds to the $\lambda_W=-1$ polarization, whereas a signal in the $\theta\in[\pi - \epsilon, \pi]$ interval characterizes the $\lambda_W=+1$ polarization. It is clear that, since the $W$ boson is destroyed in the decay process, the corresponding measurement procedure cannot be described as a projective measurement; rather, it should be interpreted as a POVM element. While various choices for the POVM set are possible, we shall adopt a specific set suitable for the purposes of this paper. We define the following operators, in terms of which the POVM set will be constructed:
\begin{eqnarray}
 \label{WbosonPOVM1}
 \left(\Gamma^{\Theta_j}_{s_W}\right)_{\nu,\nu^\prime}=\frac{|\vec k_1|_{\text{CM}}}{16\pi m_W^2}\int_0^\pi \left(M^{\Theta_j}_{s_W}\right)^\dagger_\nu \left(M^{\Theta_j}_{s_W}\right)_{\nu^\prime} \sin \theta d\theta \\
 \label{WbosonPOVM2}
 \left(M^{\Theta_j}_{s_W}\right)_\nu=\frac{g_W}{\sqrt 2}\bar u(k_1,s_1)\gamma_\mu \hat L v(k_2,s_2) {\Lambda^\mu}_\nu \;\chi_{\Theta_j}(\theta).
\end{eqnarray}
Here, ${\Lambda^\mu}_\nu$ denotes the Lorentz transformation matrix from the rest frame of the $W$ boson to the lab frame, defined through $\varepsilon^\mu(k,\lambda_W)={\Lambda^\mu}_\nu (\varepsilon^\nu_{\lambda_W})_{\mathrm{RF}}$, and $\chi_{\Theta_j}(\theta)$ represents the indicator function, i.e., $\chi_{\Theta_j}(\theta)= \{ \begin{smallmatrix} 1 & ;\;\; \theta \in \Theta_j \\ 0 & ;\;\; \theta \notin \Theta_j   \end{smallmatrix}$. The set $\left\{ \Theta_j \right\}_{j=1}^n$ forms a partition of the interval $[0,\pi]$ that is, $\bigcup_{j=1}^n \Theta_j=[0,\pi]$ and $\Theta_j \cap \Theta_{j'}=\emptyset$ for $j\neq j'$. In the derivation of Eq. (\ref{WbosonPOVM1}), the decay is evaluated in the $W$ boson's rest frame (CM frame), where $\theta$ denotes the polar angle of the lepton. In this frame, the Lorentz transformation reduces to the identity, i.e., ${\Lambda^\mu}_\nu=\delta^\mu_\nu$. During the integration azimuthal symmetry is assumed, and the $\phi$ integration has been performed. Observe that the rest frame polarization vector $(\varepsilon^\nu_{\lambda_W})_{\rm RF}$ of the incoming $W$ boson has been factored out in the amplitude factor $(M^{\Theta_j}_{s_W})_\nu$ defined in Eq.~(\ref{WbosonPOVM2}). Therefore, $M^{\Theta_j}_{s_W}:\mathcal{H}_{s_W}\rightarrow\mathbb{C}$ defines a linear functional on the $W$ boson spin Hilbert space, whereas the quantities $\Gamma^{\Theta_j}_{s_W}$ defined in (\ref{WbosonPOVM1}) act as operators on the same space, $\Gamma^{\Theta_j}_{s_W}: \mathcal{H}_{s_W} \to \mathcal{H}_{s_W}$. Let $\{|s_W\rangle_\text{in}\}_{s_W}$ denote a basis of the Hilbert space $\mathcal{H}_{s_W}$ written in Dirac notation. These basis states represent the physical polarization states of the incoming $W$ boson and are identified with the corresponding rest frame polarization vectors $(\varepsilon^\nu_{\lambda_W})_{\rm RF}$. The POVM elements $\{E_j\}_{j=1}^n$ are constructed from the diagonal matrix elements of $\Gamma^{\Theta_j}_{s_W}$ in this basis according to
\begin{eqnarray}
 \label{WbosonPOVM3}
 E_j=\sum_{s_W}\frac{{}_{\text{in}}\langle s_W| \Gamma^{\Theta_j}_{s_W}|s_W\rangle_\text{in}}{\Gamma(s_W)}\; |s_W \rangle_{\text{in}}\; {}_{\text{in}}\langle s_W|
\end{eqnarray}
,where
\begin{eqnarray}
 \label{WbosonPOVM4}
 \Gamma(s_W)=\sum_{j=1}^n {}_{\text{in}}\langle s_W| \Gamma^{\Theta_j}_{s_W}|s_W\rangle_\text{in}.
\end{eqnarray} 
 Note that the number of elements in the POVM set can be greater than $\dim \mathcal{H_{S_W}}$, i.e., $n \ge 3$. If the initial $W$ spin state is in the helicity basis, then we set $s_W = \lambda_W = +1, -1, 0$ in Eq. (\ref{WbosonPOVM3}). In this case, the coefficients preceding the spin projections are derived from the helicity amplitudes [Eqs. (\ref{Wdecayhelicityamplitudes1})–(\ref{Wdecayhelicityamplitudes6})] by imposing constraints on the corresponding angular regions. A given element $E_j$ encodes information regarding the $W$ polarization for bosons decaying into the angular region $\Theta_j$. If the $W$ boson is prepared in the polarization state $\lambda_W=x$, $(x=+1,-1,0)$ the decay probability into this angular region is given by
\begin{eqnarray}
\label{WPOVMProb1}
P_j(\lambda_W=x)=\langle \lambda_W=x|E_j|\lambda_W=x\rangle,\;\;\;x=+1,-1,0.
\end{eqnarray}
Here, the normalization is performed with respect to the leptonic decay channel of the $W$ boson, i.e., the probabilities are conditioned on leptonic decays of the $W$ and therefore sum to unity when integrated over the full angular range within this channel. The chosen POVM elements cannot always distinguish the polarization of the $W$ boson. However, elements associated with certain angular regions can, for some measurement outcomes, isolate a single polarization state by excluding the other two. For sufficiently small $\epsilon$, consider $E_1$ and $E_2$ corresponding to the angular regions $\Theta_1 = [0,\epsilon]$ and $\Theta_2 = [\pi - \epsilon, \pi]$. As seen in Figure \ref{fig2}, $P_1(\lambda_W = +1, 0) \le \delta$ and $P_1(\lambda_W = -1) > \delta$, with $\delta \to 0$ as $\epsilon \to 0$. In this limit, whenever $E_1$ yields an outcome, the polarization $\lambda_W = -1$ is determined with certainty. Similarly, we have $P_2(\lambda_W = -1, 0) \le \delta$ and $P_2(\lambda_W = +1) > \delta$, implying that whenever $E_2$ yields an outcome, the polarization $\lambda_W = +1$ is determined with certainty. For qutrits, several classes of $\psi$-epistemic models --namely, those satisfying the ontic indifference condition, $\delta$-continuous models with $\delta > 1 - \sqrt{2/3}$, and symmetric as well as strongly symmetric maximally nontrivial models-- can yield predictions that deviate from the zero-probability predictions of QM, as discussed in Section 2. Within these classes, $\psi$-epistemic models can yield outcomes in conflict with QM by assigning probabilities $P_1(\lambda_W = +1, 0) > \delta$ or $P_2(\lambda_W = -1, 0) > \delta$. This gives rise to a charged-lepton signal near $\theta \approx 0$ or $\theta \approx \pi$ that is not predicted by QFT. The presence of such a signal may be interpreted as evidence for the existence of $\psi$-epistemic models, whereas its absence places constraints on the model, analogous to Eq. (\ref{Poisson}), via a Poisson analysis. Moreover, any deviation from the zero-probability predictions of QM necessarily reduces the probability of decay into other angular regions due to normalization. This, in turn, leads to a modification of the angular distribution of the differential decay rate. Such a modification constitutes a testable signal, for instance, via a $\chi^2$ fit to the angular distribution.

Testing $\psi$-epistemic model predictions in particle physics experiments requires addressing the preparation of vector bosons in definite polarization states. As in the case of decay processes, the angular distributions of $W$ and $Z$ bosons are correlated with their polarizations in various weak production channels. Selecting bosons produced in certain angular regions provides a way to prepare specific polarization states.\footnote{This approach represents a probabilistic preparation scheme. In an actual experimental setup, the purity of the prepared states would depend on specific factors such as detector resolution, background processes, and the choice of kinematic cuts.} To provide a simple example, let us consider the production process $\bar u d \to W^- H$. This process involves three Feynman diagrams at tree level. However, the $uuH$ and $ddH$ couplings are proportional to the $u$ and $d$ quark masses and are negligible at the energy scales relevant for $WH$ production. As a result, the contributions from Higgs couplings to $u$ and $d$ quarks are negligible, with the process being dominated by the diagram containing the $WWH$ vertex. The polarized Feynman amplitudes for this diagram are given in Appendix B. In Figure \ref{fig3}, surface plots of the differential cross section are presented as functions of the $W$ boson scattering angle $\theta$ and the polar angle $\theta^*$ of its polarization vector (see Appendix B for details). As evident from the plots, the boson polarization exhibits a correlation with the scattering angle. For a fixed scattering angle, the polarized differential cross sections determine the relative probabilities of the possible polarization states. Selecting events within a specific angular region thus yields an ensemble enriched in specific polarization states, effectively realizing a probabilistic preparation. This allows one to control the polarization content of the ensemble through kinematic selection. As discussed previously, the polarization of the produced $W$ bosons is determined from the angular distributions of their decay products observed in detectors, with the analysis restricted to events passing appropriate kinematic selections. Regarding the treatment of the preparation and measurement processes together, the following remarks are in order. We employ two coordinate systems. The first is the partonic center-of-mass frame (coordinate system $(x,y,z)$) of the process $\bar{u} d \to W H$, where the $z$-axis is chosen along the incoming $d$ quark direction and the scattering plane defines the $x\!-\!z$ plane; in this frame, $\theta$ denotes the polar scattering angle of the $W$ boson. The second system is the rest frame of the produced $W$ boson (coordinate system $(x^\prime,y^\prime,z^\prime)$), where the decay amplitudes are expressed. The $(x^\prime,y^\prime,z^\prime)$ coordinate system is obtained from the $(x,y,z)$ system by a Lorentz boost along the direction $\vec{p}_W/|\vec{p}_W|$. Here, $\vec{p}_W$ denotes the three-momentum of the $W$ boson in the $(x,y,z)$ frame. In the $(x^\prime,y^\prime,z^\prime)$ coordinate system, the $W$ boson is at rest and helicity basis is therefore not uniquely defined. Instead, a spin-quantization axis, defined by the polarization vector $\vec{\varepsilon}^{\,\prime}_{3}$, is introduced. The angles $\theta^*$ and $\phi^*$ denote the polar and azimuthal angles of $\vec{\varepsilon}^{\,\prime}_{3}$ in the $(x^\prime,y^\prime,z^\prime)$ system. For each event of the full process $\bar u d \to W^- H\to \ell \bar{\nu}_\ell H$, we choose the angles $\theta^*$ and $\phi^*$ such that the vector $\vec{\varepsilon}^{\,\prime}_{3}$ lies in the decay plane of the $W$ boson. In this case, the helicity amplitudes given in Eqs. (\ref{Wdecayhelicityamplitudes1})--(\ref{Wdecayhelicityamplitudes6}) are evaluated with $\theta$ identified as the angle between the charged lepton momentum $\vec p_\ell$ and $\vec{\varepsilon}^{\,\prime}_{3}$. Here $\theta$ should not to be confused with the $W$ boson scattering angle introduced earlier in the partonic center-of-mass frame.

\section{Conclusions and Discussion}

In this work, we have shown that certain classes of $\psi$-epistemic models can give rise to observable consequences in particle physics. We illustrated these consequences explicitly in the contexts of neutrino phenomenology and polarized gauge boson decay processes. The results demonstrated in this work through several representative processes open the way to numerous further studies in particle phenomenology. The suppression of SM neutrino event rates predicted by $\psi$-epistemic models may also appear in various beyond the SM scenarios, such as active–sterile neutrino mixing \cite{MINOS,NOvA}, neutrino decays into invisible channels \cite{Gonzalez-Garcia,Ternes}, and new neutrino–matter interactions \cite{Antusch,Miranda}. Similarly, new interactions that modify the weak interaction structure of gauge bosons may alter their angular distributions and thereby produce signals similar to those generated by $\psi$-epistemic models. Therefore, in the experimental investigation of such beyond the SM scenarios, one should also take into account the possibility that an observed signal may originate not from a new particle physics model or interaction, but from a modification of quantum theory itself.

The results presented in this paper do not rely on any particular relativistic or covariant formulation of ontological models. The essential assumption is merely that measurements implemented through relativistic processes can still be represented within the ontological framework by well-defined response functions and probabilities. Indeed, it should be stressed that relativistic considerations play only a secondary role in our argument. Their primary purpose is to provide a relativistic framework for the measurement procedures rather than to generate the deviation from the predictions of QM. To be more precise, the POVM elements constructed for the representative processes considered in this work are defined on the spin Hilbert spaces of non-relativistic QM. Accordingly, these POVM elements can be regarded as measurements acting on non-relativistic quantum states. The relativistic and QFT ingredients enter only through the operational realization of these measurements, which relies on QFT observables such as scattering cross sections and decay rates. The no-go theorems discussed in section 2 do not rely on any specific dynamical realization of the measurement procedure. In these proofs, no special assumption is made regarding the physical nature of the measurement $M$; only the normalization condition (\ref{normalization}) for the response functions $\xi_{k,M}(\lambda)$ is required. Therefore, whether a measurement is implemented through a conventional non-relativistic apparatus or through a process described by QFT, such as a scattering or a decay, does not affect the logical structure of the proof, provided that the corresponding response functions $\xi_{k,M}(\lambda)$ are well defined and satisfy the normalization condition.

Ontological models have been developed primarily within the framework of non-relativistic QM. However, since relativistic phenomena are an essential part of nature, a complete assessment of the physical viability of such models ultimately requires understanding their status in relativistic settings. At first sight, the assumption that an ontological model respects Lorentz symmetry may appear too restrictive in view of the difficulties encountered by several well-known realist interpretations of QM. However, these difficulties are most naturally associated with $\psi$-ontic models. Harrigan and Spekkens showed that $\psi$-ontic ontological models reproducing the quantum statistics cannot be local \cite{Harrigan}. Although nonlocal ontological models are not necessarily incompatible with relativity,\footnote{Notable examples include relativistic spontaneous collapse models.} they are often in tension with it.  Well-known examples in this regard include the de Broglie-Bohm pilot-wave theory \cite{Bohm1,Bohm2} and the original GRW spontaneous collapse model \cite{Ghirardi}. Another important result pointing to the tension between $\psi$-ontic models and relativity is Hardy's paradox \cite{Hardy:1992zz}. Hardy argued that realistic interpretations of QM face serious difficulties in maintaining Lorentz invariance and suggested that introducing a preferred reference frame provides a way out of these difficulties. Unlike $\psi$-ontic models, there is no no-go theorem requiring $\psi$-epistemic models to be nonlocal. Furthermore, the difficulties highlighted by Hardy's paradox arise from assigning measurement independent ontic reality to the quantum state of a system. Consequently, $\psi$-epistemic models, which attribute an epistemic rather than ontic status to the quantum state, are largely shielded from Hardy's criticism. This suggests that these models offer greater flexibility than $\psi$-ontic models in the development of an ontology compatible with relativity.

\newpage
\appendix
\section*{Appendix A: Helicity Amplitudes For $W\to \ell \bar{\nu}_\ell$ Decay}

\begin{figure}[h]
\centering
\includegraphics[scale=0.8]{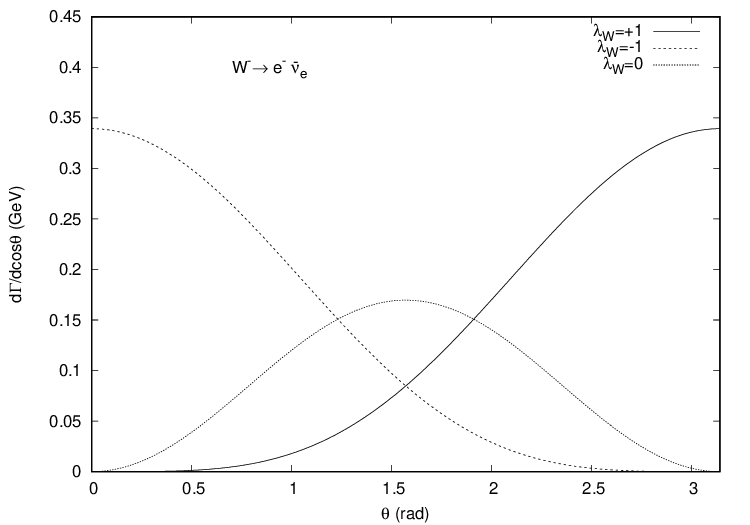}
\caption{Angular distribution of the polarized differential decay rate $\frac{d\Gamma(\lambda_W)}{d\cos \theta}$ for the decay $W^-\to e^- \bar{\nu}_e$. The final-state electron is unpolarized. \label{fig2}}
\end{figure}

This appendix presents the helicity amplitudes and the decay rate for the $W^-\to \ell \bar{\nu}_\ell$ decay, where $\ell=e^-, \mu^-, \tau^-$ represents the charged leptons. In the $W$ boson rest frame, the nonvanishing helicity amplitudes $M(\lambda_\ell,\lambda_{\bar \nu_\ell},\lambda_W)$ for this decay are given by
\begin{eqnarray}
\label{Wdecayhelicityamplitudes1}
&&M(+,+,0)=\frac{g_W}{\sqrt 2}\;\frac{m_\ell}{m_W}\sqrt{m_W^2-m_\ell^2}\;\cos\theta \\
\label{Wdecayhelicityamplitudes2}
&&M(+,+,+)=-\frac{g_W}{2}\;\frac{m_\ell}{m_W}\sqrt{m_W^2-m_\ell^2}\;\sin\theta \\
\label{Wdecayhelicityamplitudes3}
&&M(+,+,-)=\frac{g_W}{2}\;\frac{m_\ell}{m_W}\sqrt{m_W^2-m_\ell^2}\;\sin\theta \\
\label{Wdecayhelicityamplitudes4}
&&M(-,+,0)=-\frac{g_W}{\sqrt 2}\;\sqrt{m_W^2-m_\ell^2}\;\sin\theta \\
\label{Wdecayhelicityamplitudes5}
&&M(-,+,+)=\frac{g_W}{2}\;\sqrt{m_W^2-m_\ell^2}\;(1-\cos \theta) \\
\label{Wdecayhelicityamplitudes6}
&&M(-,+,-)=\frac{g_W}{2}\;\sqrt{m_W^2-m_\ell^2}\;(1+\cos \theta)
\end{eqnarray}
,where $\theta$ is the polar angle of the charged lepton, $m_W$ and $m_\ell$ are the masses of the $W$ boson and the charged lepton respectively and $g_W=\sqrt{4\pi\alpha}/\sin \theta_W$. Here, $\alpha$ denotes the fine-structure constant and $\theta_W$ the Weinberg angle. In computing the helicity amplitudes, the neutrino mass is neglected in view of $m_W\gg m_\nu$; consequently, only the amplitudes corresponding to a positive-helicity antineutrino are nonvanishing. In the $W$ boson rest frame, the differential decay rate is given by
\begin{eqnarray}
\label{difdecayrate}
\frac{d\Gamma}{d\cos \theta}=\frac{(m_W^2-m_\ell^2)}{32\pi m_W^3}\;|M|^2.
\end{eqnarray}
If the polarization of the final-state charged lepton is not observed, the lepton is treated as unpolarized, and the decay rate is computed using $|M(\lambda_W)|^2=\sum_{\lambda_\ell}|M(\lambda_\ell,+,\lambda_W)|^2$. Figure \ref{fig2} shows the angular distribution of the polarized differential decay rate $\frac{d\Gamma(\lambda_W)}{d\cos \theta}$ for unpolarized final-state electrons. Similar distributions can also be obtained for $\ell=\mu,\tau$.

\appendix
\newpage
\section*{Appendix B: Polarized Amplitudes For $\bar u d \to W^- H$}

\begin{figure}[h]
\centering

\begin{minipage}{0.32\textwidth}
\centering
\includegraphics[width=\linewidth]{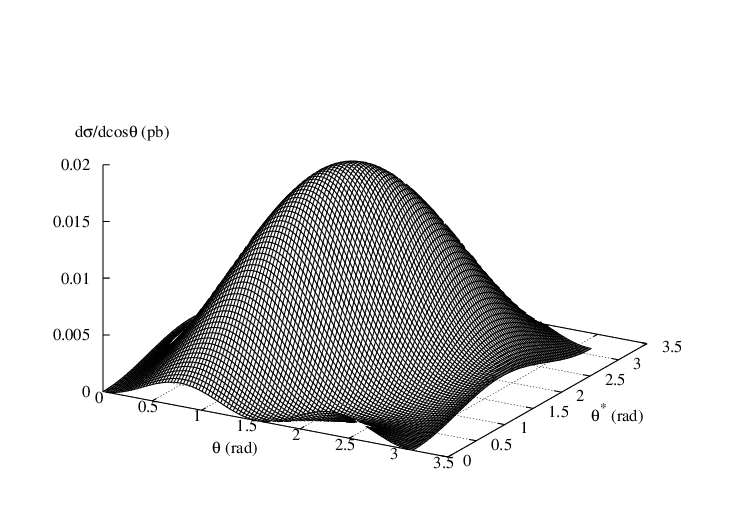}
\end{minipage}\hfill
\begin{minipage}{0.32\textwidth}
\centering
\includegraphics[width=\linewidth]{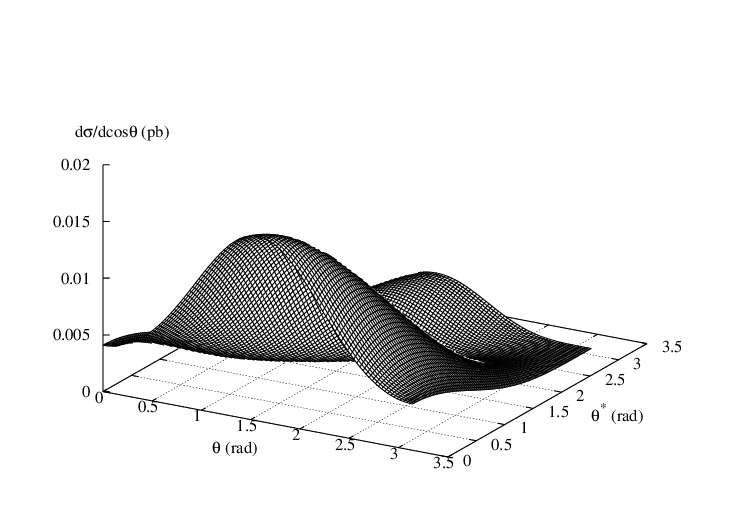}
\end{minipage}\hfill
\begin{minipage}{0.32\textwidth}
\centering
\includegraphics[width=\linewidth]{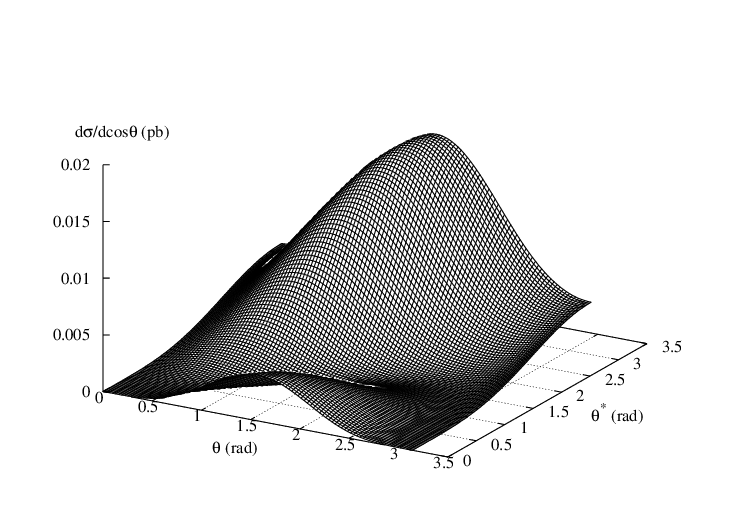}
\end{minipage}

\caption{Surface plots of the differential cross section for the $\bar{u} d \to W^- H$ process as a function of the parameters $\theta$ and $\theta ^{*}$ at $\sqrt{s} = 500$ GeV. The plots from left to right correspond to the W boson polarization states $\lambda_W = 0$, $\lambda_W = +1$, and $\lambda_W = -1$, respectively, where the initial-state quarks are unpolarized. The polarization azimuthal angle is fixed at $\phi^* = 0$.\label{fig3}}
\end{figure}

The leading-order Feynman amplitude for this process is given by
\begin{eqnarray}
\label{amplitudeWHproduction}
M=\frac{g_W^2 m_W V_{ud}}{\sqrt 2(q^2-m_w^2)}\bar v(p_2,s_2)\gamma_\mu \hat L u(p_1,s_1)(g^{\mu\nu}-\frac{q^\mu q^\nu}{m_w^2})\varepsilon^{\ast}_\nu(k,\lambda_W)
\end{eqnarray}
, where $p_{1}$ and $p_{2}$ denote the momenta of the $d$ and $u$ quarks, with $q = p_1 + p_2$, and $V_{ud}$ representing the corresponding CKM matrix element. The nonvanishing polarized amplitudes $M(\lambda_d,\lambda_{\bar u},\lambda_W)$ in the center-of-mass (CM) frame are given as follows:
\begin{eqnarray}
\label{WHpolamplitudes1}
M(-,+,0)&&=\frac{g_W^2 V_{ud}\sqrt{s}}{2\sqrt{2}(s-m_W^2)}
\Bigg[
(E_W - m_W)\cos\theta \cos\theta^{*}\sin\theta \nonumber \\
&&+ \Big(
- i\, m_W\sin\phi^{*}
+ \cos\phi^{*} \big(m_W + (E_W - m_W)\sin^2\theta \big)
\Big)\sin\theta^{*}
\Bigg] \\
\label{WHpolamplitudes2}
M(-,+,+)&&=\frac{g_W^2 V_{ud}\sqrt{s}}{8(s-m_W^2)}
\Bigg[
i\Big(
E_W + m_W + (-E_W + m_W)\cos(2\theta) + 2 m_W \cos\theta^{*}
\Big)\sin\phi^{*} \nonumber \\
&&
+\, 2\cos\phi^{*}\Big(
- m_W(1 + \cos\theta^{*})
+ (-E_W + m_W)\cos\theta^{*}\sin^2\theta
\Big) \nonumber \\
&&
+\, (E_W - m_W)\sin(2\theta)\sin\theta^{*}
\Bigg] \\
\label{WHpolamplitudes3}
M(-,+,-)&&=\frac{g_W^2 V_{ud}\sqrt{s}}{4(s-m_W^2)}
\Bigg[
i\,\sin\phi^{*}\Big(
m_W(-1 + \cos\theta^{*})
+ (-E_W + m_W)\sin^2\theta
\Big) \nonumber \\
&&
+\, \cos\phi^{*}\Big(
m_W
+ \cos\theta^{*}\big(
- m_W + (-E_W + m_W)\sin^2\theta
\big)
\Big) \nonumber \\
&&
+\, (E_W - m_W)\cos\theta \sin\theta \sin\theta^{*}
\Bigg]
\end{eqnarray}
Here, $s$ is the Mandelstam parameter and $E_W=\frac{s+m_W^2-m_H^2}{2\sqrt s}$. $\theta$ denotes the polar scattering angle of the $W$ boson, while $\theta^*$ and $\phi^*$ denote the polar and azimuthal angles of the polarization vector $\vec{\varepsilon}^{\,\prime}_{3}$ in the rest frame of the $W$ boson, as defined in Eqs.~(\ref{spin1rest1})--(\ref{spin1rest3}). Figure \ref{fig3} presents the surface plots of the differential cross section as a function of the parameters $\theta$ and $\theta^{\ast}$. In the plots, the azimuthal angle of the polarization vector is taken to be $\phi^*=0$. The $u$ and $d$ quark masses are neglected throughout the calculations.

\end{document}